\documentclass{aa}  

\usepackage{txfonts}
\usepackage{hyperref}      
\hypersetup{colorlinks=true,linkcolor=blue,citecolor=blue,filecolor=blue,urlcolor=blue,}
\usepackage{placeins}    
\usepackage{graphicx}   

\makeatletter
\renewcommand*\aa@pageof{, page \thepage{} of \pageref*{LastPage}}
\makeatother

\usepackage[version=4]{mhchem}
\usepackage{color}

\begin{document} 

   \title{CO\textsubscript{2}-rich protoplanetary discs as a probe of dust radial drift and trapping}

   \author{Andrew D. Sellek
          \inst{1}
          \and
          Marissa Vlasblom
          \inst{1}
          \and
          Ewine F. van Dishoeck
          \inst{1,2}
          }

   \institute{Leiden Observatory, Leiden University, 2300 RA Leiden, The Netherlands\\
   \email{sellek@strw.leidenuniv.nl}
   \and
   Max-Planck Institut f\"{u}r Extraterrestrische Physik (MPE), Giessenbachstr. 1, 85748, Garching, Germany}

   \date{Received 16 June 2024 / Accepted 15 November 2024}

 
  \abstract
   {Mid-infrared spectra indicate considerable chemical diversity in the inner regions of protoplanetary discs, with some being \ce{H2O}-dominated and others \ce{CO2}-dominated. Sublimating ices from radially drifting dust grains are often invoked to explain some of this diversity, particularly with regards to \ce{H2O}-rich discs.}
   {We model the contribution made by radially drifting dust grains to the chemical diversity of the inner regions of protoplanetary discs. These grains transport ices ---including those of \ce{H2O} and \ce{CO2}--- inwards to snow lines, thus redistributing the molecular content of the disc. As radial drift can be impeded by dust trapping in pressure maxima, we also explore the difference between smooth discs and those with dust traps due to gas gaps, quantifying the effects of gap location and formation time.}
   {We used a 1D protoplanetary disc evolution code to model the chemical evolution of the inner disc resulting from gas viscous evolution and dust radial drift. We post-processed these models to produce synthetic spectra, which we analyse with 0D LTE slab models to understand how this evolution may be expressed observationally.}
   {Discs evolve through an initial \ce{H2O}-rich phase as a result of sublimating ices, followed by a \ce{CO2}-rich phase as \ce{H2O} vapour is advected onto the star and \ce{CO2} is advected into the inner disc from its snow line. The introduction of traps hastens the transition between the phases, temporarily raising the \ce{CO2}/\ce{H2O} ratio.
   However, whether or not this evolution can be traced in observations depends on the contribution of dust grains to the optical depth. If the dust grains become coupled to the gas after crossing the \ce{H2O} snow line ---for example if bare grains fragment more easily than icy grains--- then the dust that delivers the \ce{H2O} adds to the continuum optical depth and obscures the \ce{H2O}, preventing any evolution in its visible column density.
   However, the \ce{CO2}/\ce{H2O} visible column density ratio is only weakly sensitive to assumptions about the dust continuum obscuration, making it a more suitable tracer of the impact of transport on chemistry than either individual column density. This can be investigated with spectra that show weak features that probe deep enough into the disc.
   The least effective gaps are those that open close to the star on timescales competitive with dust growth and drift as they block too much \ce{CO2}; gaps opened later or further out lead to higher \ce{CO2}/\ce{H2O}. This leads to a potential correlation between \ce{CO2}/\ce{H2O} and gap location that occurs on million-year timescales for fiducial parameters.
   }
   {Radial drift, especially when combined with dust trapping, produces \ce{CO2}-rich discs on timescales longer than the viscous timescale at the \ce{H2O} snow line (while creating \ce{H2O}-rich discs at earlier times). Population analyses of the relationship between observed inner disc spectra and large-scale disc structure are needed to test the predicted role of traps.}


   \maketitle
%

\section{Introduction}
The formation and growth of planets starts around a young star during the protoplanetary disc phase.
Although the most direct observations of protoplanets so far \citep{Keppler_2018,Haffert_2019,Mesa_2019,Currie_2022,Hammond_2023} are of (super-)Jupiter-mass protoplanets orbiting at tens of astronomical units (au), the inner few au are also of special interest, being the location where inward pebble drift leads to a higher midplane dust-to-gas ratio \citep{Weidenschilling_1977}, which may be conducive to planetesimal formation \citep{Drazkowska_2016}. The inner disc is also the site of the water snow line ---where dust at the disc midplane becomes hot enough for \ce{H2O} to thermally desorb and enter the gas phase---, which may assist the concentration of these building blocks and subsequent planet formation just outside the snow line 
\citep{Stevenson_1988,Ida_2016,Schoonenberg_2017,Drazkowska_2017,Schoonenberg_2018}.
More massive planets may also potentially form in the inner disc, or otherwise form at larger radii, where substructures often attributed to planets are commonly observed \citep{Andrews_2020,Bae_2023}, and migrate inwards \citep{Lodato_2019}.
As protoplanets will inherit their compositions from the disc material, information on the chemistry of the inner disc region sheds light on the initial chemical abundances of planets that form in, or migrate to, the inner disc.

These inner regions are warm enough to excite line emission at the mid-infrared (MIR) wavelengths, which has allowed the volatile molecular content of this planet-forming material to be studied with several generations of high-resolution spectrographs and space-based telescopes \citep{Pontoppidan_2014}. JWST is now providing an unprecedented view of this gas-phase chemistry \citep{vanDishoeck_2023,Kamp_2023,Henning_2024}, and its intriguing diversity across the disc population. For example, the multitudinous abundant hydrocarbons and weak or absent water seen towards very low-mass stars \citep{Tabone_2023,Arabhavi_2024} confirms the carbon-rich nature of their discs previously inferred from abundant \ce{C2H2} in Spitzer spectra \citep{Pascucci_2009b,Pascucci_2013}.
On the other hand, in T Tauri stars with masses of $\gtrsim 0.3\,M_{\sun}$, the main molecules tend to be \ce{H2O} and \ce{CO2}. 
In this case, diversity is seen between `\ce{CO2}-dominated spectra', such as those of GW Lup \citep{Grant_2023} and CX Tau \citep{Vlasblom_2024b} in which the Q-branch of \ce{CO2} is particularly prominent \citep{Pontoppidan_2010} ---and in which the \ce{^13CO2} isotopologue may also be detected---, and spectra where the forest of water lines is the most prominent, such as those of CI Tau, GK Tau, HP Tau, IQ Tau \citep{Banzatti_2023}, Sz 98 \citep{Gasman_2023}, SY Cha \citep{Schwarz_2024}, and DR Tau \citep{Temmink_2024a,Temmink_2024b}.

To understand the origin of this diversity, we must identify the factors that determine the inner disc chemistry: for example, we must decipher whether this chemistry reflects the chemical equilibrium, given the temperature, stellar radiation field, dust distribution, and cosmic ray flux \citep[e.g.][]{Eistrup_2016,Eistrup_2018}, or a non-equilibrium maintained by significant transport of volatiles, either in the gas phase following the accretion flow or as ices on inwardly drifting dust grains. Moreover, the extent to which the observed chemistry relies on which of the layers of the disc are probed by the measurements remains uncertain.

To see how chemistry and transport compete, \citet{Booth_2019} compared a static disc model to models with purely viscous evolution and then also to models incorporating dust radial drift \citep[see also][]{CevallosSoto_2022}.
For example, for \ce{CO2} or \ce{NH3}, \citet{Booth_2019} find that the timescale for gas-phase viscous transport inwards from their snow lines is shorter than the timescale on which the bulk of the material at the midplane becomes transformed by gas-phase chemical reactions in the inner disc. This allows transport to entirely counteract any decrease in their abundances in the inner disc, which is agreement with the results of \citet{Bosman_2018a}, who find that highly enhanced destruction rates would be needed to significantly depress inner disc \ce{CO2} abundances.
On the other hand, \cite{Booth_2019} also find that for a more volatile molecule such as \ce{CH4}, which is transformed 
(by conversion into larger hydrocarbons) on a shorter timescale relative to its viscous transport time \citep[given the low typical viscosities inferred at tens of au; e.g.][]{Rosotti_2023}, viscous transport is unable to erase the signature of chemical reactions.
Radial drift further enhances the speed of the inward transport of the volatiles. This in turn further increases the degree to which major volatiles such as \ce{H2O} and \ce{CO2} are enhanced, and negligible differences are seen in their profiles between models with and without chemical reactions. The strong \ce{CH4} abundance decrease in the inner disc is now prevented by replacement by drifting ices, despite an efficient conversion to hydrocarbons when reactions are included.

Consequently, disc evolution models have been used to explore the evolution of the C/H, O/H, and C/O ratios in the inner disc due to the successive enhancement of \ce{H2O}, \ce{CO2}, and then finally hydrocarbons (such as \ce{CH4}) and CO by drifting pebbles \citep[e.g.][]{Booth_2017}. This happens because the ices of these molecules have increasingly large binding energies and therefore have snow lines at increasingly large radii. This means that these molecules are deposited by desorption from drifting pebbles successively further from the star, and then take longer to be advected into the inner disc by the gas.
However, the absolute timing of these phases may vary, for example as a function of stellar mass, as lower-mass stars have correspondingly cooler discs, closer-in snow lines, and shorter timescales for transport \citep{Mah_2023}.

However, most studies which investigate these phenomena neglect the fact that the aforementioned substructures are frequently observed and are typically understood to represent dust traps that would interrupt the supply of volatiles and could solve several issues in planet formation and disc evolution modelling \citep{Birnstiel_2024}.
While focused on photo-evaporation, \citet{Lienert_2024} also include a single example of a planet-carved gap in a model of disc chemical evolution by transport, but do not systematically study the effect of the gap's properties.
A more systematic exploration ---albeit only in relation to \ce{H2O}--- was conducted by \citet{Kalyaan_2021,Kalyaan_2023}, who demonstrated the impact of dust traps at different radii on the \ce{H2O} vapour mass in the inner disc. A key finding was that, the closer the trap to the star, the greater the fraction of the \ce{H2O} ice reservoir it could isolate from reaching the inner disc and hence the more the enhancement of the \ce{H2O} vapour mass due to drift is suppressed.
Moreover, \citet{Kalyaan_2021} argue that drift simultaneously shrinks dust discs and delivers water to the inner disc, and may thus explain the anti-correlation between the luminosity of mid-infrared (MIR) water lines and the radius containing 90\% of the continuum flux at millimetre wavelengths \citep{Banzatti_2020}; however \citet{Kalyaan_2021} do not directly forward model such quantities. Finally, \citet{Mah_2024} also explored the role of the trap depth and formation time on the inner disc \ce{H2O} abundance and also the C/O ratio. 

In order to understand the connection between the disc structure at large radii and the diversity in the MIR molecular spectra of T Tauri discs, we built on these previous works by modelling the delivery of both \ce{CO2} and \ce{H2O} to the inner regions of discs with dust traps.
We focus on these two molecules as they are  (a)  very widely detected in spectra from JWST's Mid-Infrared Instrument Medium-Resolution Spectroscopy mode (MIRI-MRS) spectra (unlike e.g. \ce{NH3}) and are  (b) relatively insensitive to chemical processing throughout the bulk of the disc (unlike hydrocarbons such as \ce{CH4} or \ce{C2H2}). 
We conducted this investigation using a 1D disc evolution code \citep{Booth_2017}, and in Sect. \ref{sec:methodsDisc} we summarise its key features and the modifications we made to the chemistry module. To make the connection to observations more direct, we then produced simple forward models of the MIR spectra and retrieved representative column densities, temperatures, and emitting areas using 0D slab models assuming local thermodynamic equilibrium (LTE) as described in Sect. \ref{sec:methodsSpectrum}. Section \ref{sec:results} presents the results of this exercise in terms of the evolution of the vapour masses and retrieved column densities. We compare the predicted ranges to observed discs, and explore any dependence on gap location at different stages of disc evolution. In Sect. \ref{sec:discussion} we discuss other processes or structural considerations that may impact our findings and consider other possible tests of our results.
Finally, we summarise our conclusions in Sect. \ref{sec:conclusions}.

\section{Disc evolution model}
\label{sec:methodsDisc}
We used the 1D disc evolution code described in \citet{Booth_2017,Booth_2018,Sellek_2020b}. The code implements the two-population dust model of \citet{Birnstiel_2012} which approximates the dust distribution as a population of small grains that follow the gas and a population of large grains that dynamically decouple from the gas. The two populations have fixed mass fractions $f_{\rm small}$ and $f_{\rm large}=1-f_{\rm small}$ respectively.
These mass fractions are calibrated against fragmentation-coagulation simulations that account for two main barriers to growth which set the maximum grain size: a) the fragmentation of grains that collide with sufficiently high relative velocity and b) the removal of dust grains by radial drift faster than they can grow.
These suggest $f_{\rm small}=0.25$ when fragmentation limits growth, or $f_{\rm small}=0.03$ when radial drift limits growth. The fragmentation limit usually applies when dust is abundant ---so mainly while discs are still very young or where dust is concentrated--- while more mature discs tend to be drift limited.

The initial disc surface density for initial mass $M_{\rm D,0}$ and radius $R_{\rm C,0}$ was given as a \citet{Lynden-Bell_Pringle_1974} profile:
\begin{equation}
    \Sigma_{t=0}(R) = \frac{M_{\rm D,0}}{2\pi R_{\rm C,0}} \frac{1}{R} \exp\left(-\frac{R}{R_{\rm C,0}}\right)
    .
\end{equation}
The temperature profile is assumed to follow a power law:
\begin{equation}
    T = T_0 \left(\frac{R}{R_0}\right)^{-1/2}
    \label{eq:Tdisc}
    ,
\end{equation}
normalised such that the aspect ratio (scale height) at $R_0=1\,\mathrm{au}$ is $h_0=0.021$ ($H_0=0.021$ au). This corresponds to a disc temperature at 1 au of $T_0=118\,\mathrm{K}$ as assumed by \citet{Kalyaan_2021} based on the median stellar luminosity in \citet{Banzatti_2020}, and also gives aspect ratios at larger radii in line with the $M_*=1\,M_{\odot}$ radiative transfer models of \citet{Sinclair_2020}.

In the following subsections, we describe more details of our approach to modelling the viscosity, dust, and traps. For ease of comparison, the fiducial parameters, which are summarised in Table\,\ref{tab:fiducialparams}, mostly follow \citet{Kalyaan_2021,Kalyaan_2023}.

\begin{table}[ht]
    \centering
    \caption{Fixed and fiducial disc model parameters used in this work}
    \label{tab:fiducialparams}
    \setlength\tabcolsep{3pt}
    \begin{tabular}{rll}
        \hline\hline
        Parameter       &   Value   & Description \\
        \hline
        $M_{*}$         &   $1\,M_{\odot}$      & Stellar mass \\
        $M_{\rm D,0}$   &   $0.05\,M_{\odot}$   & Initial disc mass \\
        $R_{\rm C,0}$   &   $70\,\mathrm{au}$   & Initial disc size \\
        $\mu$           &   $2.34$              & Mean molecular weight \\
        $h_0$           &   $0.021$             & Aspect ratio at $1\,\mathrm{au}$ \\
        $\rho_{\rm s}$  &   $1.5\,\mathrm{g\,cm^{-3}}$      & Dust material density \\
        $\epsilon$      &   $10^{-2}$           & Initial dust-to-gas mass ratio \\
        $f_{\rm small,\,frag}$  &   $0.25$      & Small dust fraction (fragmentation limit) \\
        $f_{\rm small,\,drift}$ &   $0.03$      & Small dust fraction (drift limit) \\
        \hline
        $r_{\rm in}$    & $0.1\,\mathrm{au}$    & Inner radial boundary of model \\
        $r_{\rm out}$   & $500\,\mathrm{au}$    & Outer radial boundary of model \\
        $N_{\rm cells}$ & $1000$                & Number of radial grid cells \\
        \hline
        $A_{\rm gap}$   &   $10$                & Gas gap/perturbation depth \\
        $w_{\rm gap}$   &   $1.414 H$           & Gas gap/perturbation width \\
        $t_{\rm f}$     &   $0\,\mathrm{Myr}$   & Gas gap/perturbation growth end time \\
        \hline
    \end{tabular}
\end{table}

\subsection{Key features of gas and dust transport}
The gas evolves due to the action of an effective viscosity, for which we assumed a \citet{Shakura-Sunyaev_1973} model $\nu_{\rm eff}=\alpha C_{\rm S} H$ with constant $\alpha$. This results in a viscous velocity, which may be expressed in terms of the Keplerian velocity $v_{\rm K}$ as 
\begin{equation}
    |v_{\rm visc}| = \frac{3}{2} \alpha h^2 v_{\rm K}
    \label{eq:v_visc}
    .
\end{equation}

The dust interacts with the gas through drag; this interaction may be characterised by the Stokes number
\begin{equation}
    S\!t=\frac{\pi}{2} \frac{\rho_{\rm s}a}{\Sigma_G}
    ,
\end{equation}
where $\rho_{\rm s}$ is the dust material density and $\Sigma_G$ is the gas surface density.
This interaction results in the dust radial velocity having two components \citep[e.g.][]{Dipierro_2018}: one resulting from advection with the radial motion of the gas (see Eq. \ref{eq:v_visc}), and another from radial drift \citep{Whipple_1973} which is
\begin{equation}
    |v_{\rm rd}| = {S\!t} \, h^2 \frac{\partial \ln P}{\partial \ln R} v_{\rm K}
    \label{eq:v_drift}
    .
\end{equation}
As both $3/2$ and the logarithmic pressure gradient $\frac{\partial \ln P}{\partial \ln R}$ are typically of order unity, the parameter $S\!t/\alpha$ determines the coupling of the dust to the gas. For $S\!t/\alpha \gtrsim 1$, the radial drift term dominates, while for $S\!t/\alpha \lesssim 1$, the viscous advection term dominates.

As discussed, in the early phase of disc evolution when the dust is abundant, or likewise in traps when there is no radial drift, the Stokes number of the largest grains is limited by fragmentation \citep{Birnstiel_2012}.
As collisions are typically most important between grains of similar sizes, with similar bulk velocities, the largest relative velocities are usually the random velocities due to turbulence, which results in the following limit:
\begin{equation}
    S\!t_{\rm max} = \frac{u_{\rm f}^2}{3 \alpha_{\rm t} c_{\rm S}^2}
    \label{eq:Stfrag}
    ,
\end{equation}
where $u_f$ is the fragmentation velocity and typically it is assumed that $\alpha_{\rm t}=\alpha$. Thus, for some typical values, the corresponding coupling parameter of the large, drifting dust grains in the vicinity of the water snow line at $150\,\mathrm{K}$ becomes
\begin{equation}
    \frac{S\!t_{\rm max}}{\alpha} \sim 0.6 \left(\frac{u_{\rm f}}{1\,\mathrm{m\,s^{-1}}}\right)^{2} \left(\frac{\alpha}{10^{-3}}\right)^{-2} \left(\frac{T}{150}\right)^{-1}
    \label{eq:coupling}
    .
\end{equation}

We note, however, that if the bulk turbulence (as parametrized by $\alpha_{\rm t}$) is low, other processes can dominate the relative velocities \citep{Pinilla_2021}. For example, fragmentation can occur due to the differential drift velocity between two dust grains of different sizes \citep{Birnstiel_2012}, in which case the limit becomes
\begin{equation}
    S\!t_{\rm max} = \frac{u_{\rm f} v_{\rm K}}{\gamma c_{\rm S}^2 (1-N)}
    \label{eq:Stfragdrift}
    ,
\end{equation}
where $\gamma=\frac{\partial \ln P}{\partial \ln R}$, and $N=0.5$ is the relative Stokes number assumed for the two differentially drifting grains.

\begin{figure*}[ht]
    \centering
    \includegraphics[width=0.9\linewidth]{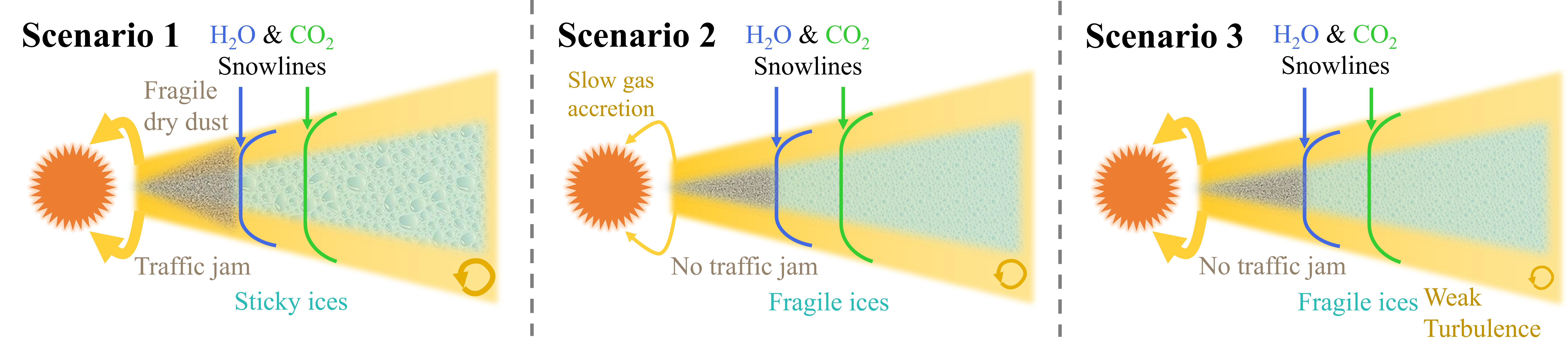}
    \caption{Cartoon depicting the three scenarios explored in this work. The thickness of the yellow arrows connecting the disc and star represents the magnitude of $\alpha$ controlling the gas evolution and accretion. The turbulence strength is indicated by the thickness and size of the circular arrow in the bottom right. The speckled region represents the dust with the brown colour indicating dry dust and the light blue colour indicating icy dust and the finer-grained pattern indicating where grains are more fragile. The vertical size of this region is inflated for illustrative purposes in locations where is a traffic jam and grains pile up.}
    \label{fig:scenarios}
\end{figure*}

\begin{figure*}[ht]
    \centering
    \includegraphics[width=0.95\linewidth]{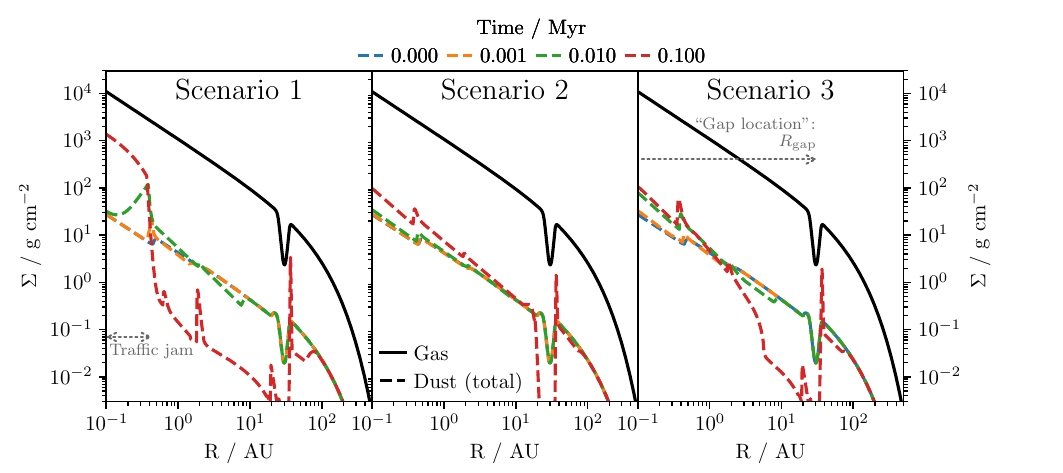}
    \caption{Example surface density profiles of gas (solid, initial value only) and (small+large) dust (dashed, at different times) for each of the three scenarios, illustrating the changing dust surface density distribution due to radial drift. The dip in the gas surface density signifies the gap, in this case at $R_{\rm gap}=30\,\mathrm{au}$ and after 0.1 Myr (red), dust has begun to accumulate in the trap at its outer edge. The traffic jam that occurs in Scenario 1 can be seen in the steep rise in the dust surface density after 0.1 Myr inside the water snow line at $\approx0.4\,\mathrm{au}$. The other small steps correspond to changes in the solid fraction at snow lines and the smaller bumps are the accumulation of ices at snow lines due to the cold finger effect.}
    \label{fig:scenariosSigma}
\end{figure*}

In this work, we considered three possible scenarios for the parameters appearing in Eqs. \ref{eq:v_visc}-\ref{eq:Stfragdrift} which we explain below and summarise in Table\,\ref{tab:scenarios} and Figs.\,\ref{fig:scenarios} and \ref{fig:scenariosSigma}. These explore different possible regimes for the turbulence and collision velocities at which fragmentation limits dust growth, two key uncertainties in current dust evolution modelling \citep{Birnstiel_2024}.
\begin{enumerate}
    \item The first parameter set assumes that icy dust grains are stickier and less fragile than dry grains, that is $u_{\rm f, dust}=1\,\mathrm{m\,s^{-1}}$ \citep{Blum_2008}, $u_{\rm f, ice}=10\,\mathrm{m\,s^{-1}}$ \citep{Gundlach_2015} and $\alpha_{\rm t}=\alpha=10^{-3}$. In this case, Eq.\,\ref{eq:coupling} predicts that dry dust grains in the inner disc that have sublimated all of their ices will be coupled to the gas, while icy dust grains at large radii will decouple. This will lead to a `traffic jam' \citep[][see also left-hand panel of Fig.\,\ref{fig:scenariosSigma}]{Saito_2011,Pinilla_2016} where the velocity of the grains decreases inside the water snow line and thus they pile up.
    \item Secondly, we considered $u_{\rm f, dust}=1\,\mathrm{m\,s^{-1}}$, $u_{\rm f, ice}=1\,\mathrm{m\,s^{-1}}$, and $\alpha=10^{-4}$. This is motivated by studies suggesting lower fragmentation velocities for icy grains \citep{Gundlach_2018,Musiolik_2019}; by also decreasing $\alpha$, we maintain the degree of coupling in the outer disc, allowing effective radial drift to continue, while also preventing the traffic jam at small radii. We note, however, that lowering $\alpha$ will also slow down the gas evolution.
    \item Finally to avoid simultaneously changing the gas evolution, we also investigated a case where $\alpha_t \neq \alpha$ \citep[c.f.][Model 6]{Pinilla_2021}: $u_{\rm f, dust}=1\,\mathrm{m\,s^{-1}}$, $u_{\rm f, ice}=1\,\mathrm{m\,s^{-1}}$, $\alpha=10^{-3}$ and $\alpha_{\rm t}=10^{-5}$ which should still allow the dust to form decoupled, drifting, pebbles both inside and outside the water snow line. In this case the fragmentation limit is due to the relative drift velocities (Eq.\,\ref{eq:Stfragdrift}).
\end{enumerate}

\begin{table*}[t]
    \centering
    \caption{Parameters relevant to the dust and gas transport that are varied between scenarios}
    \label{tab:scenarios}
    \setlength\tabcolsep{3pt}
    \begin{tabular}{rllll}
        \hline\hline
        Parameter        &   Description   & Scenario 1  & Scenario 2    & Scenario 3 \\
        \hline
        $\alpha$         &   Viscosity \citep{Shakura-Sunyaev_1973} & $10^{-3}$ & $10^{-4}$ & $10^{-3}$ \\
        $\alpha_{\rm t}$ &   Turbulence \citep{Pinilla_2021} & $10^{-3}$ & $10^{-4}$ & $10^{-5}$ \\
        $u_{\rm frag, icy}$ & Fragmentation velocity (icy grains) & $10\,\mathrm{m\,s^{-1}}$    & $1\,\mathrm{m\,s^{-1}}$ & $1\,\mathrm{m\,s^{-1}}$ \\
        $u_{\rm frag, dry}$ & Fragmentation velocity (dry grains) & $1\,\mathrm{m\,s^{-1}}$ & $1\,\mathrm{m\,s^{-1}}$ & $1\,\mathrm{m\,s^{-1}}$ \\
        \hline
        \multicolumn{2}{r}{Key Features} & Sticky ices & Fragile ices & Fragile ices \\
                        && Inner traffic jam & No traffic jam & No traffic jam \\
                        &&& Slow gas accretion & Unequal turbulent and viscous $\alpha$ \\
                        &&&& Drift-limited fragmentation \\
        \hline
    \end{tabular}
\end{table*}

\subsection{Traps}
\subsubsection{Imposition and shape of traps}
We introduced dust traps as a perturbation to the effective viscosity $\nu_{\rm eff}$ ---which governs bulk gas transport--- as follows:
\begin{equation}
    \nu_{\rm eff} = f(R,t) \alpha c_{\rm S} H
    ,
\end{equation}
where $f(R,t)$ represents the additional torques needed to create and sustain a perturbation in the gas, for example due to the action of a planet.
We note that these additional torques may not be turbulent or diffusive in nature, and so the 
radial diffusivity is assumed to be unaffected by the perturbation and as such expressed by $D=\alpha c_{\rm S} H/S\!c$ (where $S\!c\approx1$ for $S\!t\ll1$, is the Schmidt number) \citep{Stadler_2022,Kalyaan_2023,Mah_2024}.

For the spatial dependence of $f(R,t)$ we used a simple Gaussian defined in terms of the peak-to-baseline contrast $A$, the centre $R_{\rm gap}$ and the width $w_{\rm gap}$:
\begin{equation}
    f(R,t) = 1 + (A-1) g(t) \exp\left( -\frac{(R-R_{\rm gap})^2}{2w_{\rm gap}^2} \right)
    .
\end{equation}
Figure \ref{fig:scenariosSigma} includes an example of the gas surface density profile that results for $A=10$ and $R_{\rm gap}=30\,\mathrm{au}$.

Observationally, the most easily quantified parameter of any trap is its radial location in the disc, so we explored a range of gap locations: $R_{\rm gap} \in \{1,3,7,15,30,60,100\}\,\mathrm{au}$. This spans from $1\,\mathrm{au}$ - in order to explore the impact of traps that may be too close to the star to resolve with the Atacama Large Millimeter/submillimeter Array (ALMA) - to $100\,\mathrm{au}$ - just outside our fiducial value of $R_{\rm C}=70\,\mathrm{au}$. We note that the 1 au case will lie between the \ce{H2O} and \ce{CO2} snow lines for the temperature profile we have assumed.

In order to be hydrodynamically stable, the width of the trap should be at least a scale height $w_{\rm gap} \geq H(R_{\rm gap})$ \citep{Lin_1993,Dullemond_2018}. For consistency we follow \citet{Kalyaan_2021,Kalyaan_2023} and used $w_{\rm gap}=\sqrt{2}H(R_{\rm gap})$ (noting that our definition of $w_{\rm gap}$ differs from theirs by a factor $\sqrt{2}$).
To ensure that traps of this width are resolved with several grid elements requires small $\frac{\Delta R}{w_{\rm gap}}$. Since (in general) $\frac{\Delta R}{w_{\rm gap}} \leq \frac{\Delta R}{H(R_{\rm gap})} = h_0^{-1} \left(\frac{R_{\rm gap}}{R_0}\right)^{-1/4} \frac{\Delta R}{R_{\rm gap}}$, and the flaring is small (i.e. $R^{1/4}$ is a fairly weak function of R, so $\left(\frac{R_{\rm gap}}{R_0}\right)^{-1/4}\sim1$), then a sufficiently small $\frac{\Delta R}{R} < h_0$ should suffice at all radii. Hence a logarithmically spaced grid is the natural choice for this case, rather than the more computationally efficient (for solving the viscous diffusion equation) $\Delta R \propto R^{1/2}$ which provides insufficient resolution at small radii and was thus found in testing to make close-in traps artificially leaky.

Previous work by \citet{Stadler_2022} suggests that $A \gtrsim 3$ is needed to effectively trap pebbles, though somewhat lower values will still perturb the dust density profile. We used a fiducial $A=10$ in order ensure we are comfortably in this regime, without creating very steep gradients that slow down the computation.  
The gap depth is commonly related to the equivalent mass of planet that would open that gap \citep[e.g.][]{Kanagawa_2017}.
However, some caution is needed about making such interpretations. Firstly, the \citet{Kanagawa_2017} parametrization represents the viscous criterion, namely how quickly the planet can clear the gap relative to the viscous process that act to refill it; however, an alternative limit is the thermal criterion \citep{Lin_1993}, whereby the Roche radius of the planet should exceed the disc gas scale height.

Moreover, in models that solve the full \citet{Smoluchowski_1916} equation for coagulation \citep[e.g. \textsc{dustpy}][]{Stammler_2022}, dust traps are known to be leaky \citep{Drazkowska_2019,Stammler_2023}: the small grains produced by collisional fragmentation in the trap are dragged with the gas - or more importantly diffuse - across the gap. Although we do not model the production of small grains by fragmentation in detail, as we employ a fixed $f_{\rm small}$, our traps will still have a population of small dust that follows the concentration of the large dust grains. Thus the two-population model will produce diffusive leaking through the gap with a flux proportional to the diffusivity and the concentration gradient of the small dust:
\begin{equation}
    F_{\rm leak} = 2 \pi R_{\rm gap} D \Sigma \frac{\partial \epsilon_{\rm small}}{\partial R}
    \label{eq:Fleak}
    .
\end{equation}
However, by construction, the fixed $f_{\rm small}$ of the two-population model does not account for this diffusive flux in determining how much of the dust is `small' and susceptible to leaking and therefore our models will underpredict the proportion of the dust that is vulnerable to leaking. This caveat is not prohibitive to us modelling the transport of volatiles to the inner disc - so long as we are mindful of the fact that our gaps are potentially only producing the dust fluxes expected from a somewhat deeper gap. Consequently we refrain from reading too much into how the depth affects our results, or trying to equate our gap depth to a certain planet mass.

Finally, we also considered a smooth disc with no gaps at all to provide a reference model. Moreover we note that for simplicity, we do not consider discs with multiple gaps; the inner disc molecular abundances will generally be controlled by the innermost trap to the star \citep{Kalyaan_2021} as it is the final barrier isolating the dust - the inner disc is largely unaffected by whether the dust is actually trapped there or in a trap further out.

\subsubsection{Growth of traps}
One open question is at what stage of protoplanetary disc evolution most gaps and traps form, which is closely tied to the question of when planet formation begins. In order to see whether the effects of these structures on inner disc chemistry could be used to explore these questions, we explored a range of formation timescales as well as our fiducial case where the traps are present for all time (when we simply have $g(t)=1$).

If the traps are produced by planets, they can only emerge once the planets have grown sufficiently to perturb the disc \citep[e.g.][]{Lin_1993}.
When the additional torques represented by $f(R,t)$ have a planetary origin, they scale with the square of the planet mass $g(t) \propto M_{\rm P}(t)^2$ \citep{Lin_1979}.
Meanwhile, planetary mass increases due to its accretion from the disc, usually modelled as some mass-dependent power law defined by $\dot{M_{\rm P}} \propto M_{\rm P}^{b_{\rm grow}}$. Typical values are in the range $0.5<b_{\rm grow}<3$ depending on whether gas, pebble, or planetesimal accretion dominates and on what lengthscales the planet can accrete \citep{Drazkowska_2023}. Thus, we adopted the following solution for the time dependent term in the torque:
\begin{equation}
    g(t) = \min\left[1, \left( 1+ (1-b_{\rm grow}) \frac{t-t_{\rm f}}{t_{\rm grow, f}} \right)^{2/(1-b)}\right]
    .
\end{equation}
The free parameters in this model are the growth scaling $b_{\rm grow}$, the time at which the torque ceases to grow $t_{\rm f}$ and the final (i.e. at $t=t_{\rm f}$) growth timescale, $t_{\rm grow, f}$.

The most important aspect of the gaps is when the perturbation to the disc profile grows large enough to trap pebbles. In the case of a planet, this is then associated with the isolation of the pebble flux feeding its growth. The final stage of pebble accretion typically involves accretion from the whole height of the disc (2D regime) with the relative velocities dominated by Keplerian shear for larger-Stokes-number pebbles (Hill regime) \citep{Drazkowska_2023}. In this case, the growth scales as $b_{\rm grow}=2/3$ and can achieve a final growth timescale $t_{\rm grow, f}\sim10^4\,\mathrm{yr}$ at 1 au. Thus for our trap growth models, we fixed these values, and focussed on varying the end time $t_{\rm f}$ in the range $t_{\rm grow, f}<t_{\rm f}\leq\mathrm{few \times Myr}$ resulting in values $t_{\rm f} \in \{0.03,0.1,0.3,1.0\}\,\mathrm{Myr}$.
Substructures are highly common in Class II discs in the best-studied regions with ages $\gtrsim\,\mathrm{Myr}$, and are also sometimes seen during the embedded phase, for example in HL Tau \citep{ALMApartnership_2015} and IRS 63 \citep{Segura-Cox_2020} with ages $\lesssim 0.5\,\mathrm{Myr}$. Our range of formation times thus straddles the appearance of the earliest known substructures.

\subsection{Chemical tracers}
The code includes a number of molecules as tracers in both gaseous and solid (ice) phase (aside from the refractory grain components, which are only ever solid). The gaseous tracers are advected with the gas velocity, while the ices move with the dust grains.
For simplicity, in this work only transport (advection and diffusion) and state changes (freeze-out and desorption) are allowed to change the tracer abundances; there are no chemical reactions between species (which we validate in Appendix \ref{appendix:rates}). The only sink of a species is if it crosses the inner radial boundary of the model, at which point it is considered to have been advected onto the star.

\begin{table}[ht]
    \centering
    \caption{Elements included in our model and their abundances}
    \label{tab:elements}
    \setlength\tabcolsep{2pt}
    \begin{tabular}{llll}
        \hline\hline
        Element & \multicolumn{3}{c}{Initial abundance (with respect to H)} \\
                & Total & Volatile & Refractory \\
        \hline
        C       & $2.3 \times 10^{-4}$ & $1.4 \times 10^{-4}$ & $0.9 \times 10^{-4}$ \\
        O       & $5.3 \times 10^{-4}$ & $4.4 \times 10^{-4}$ & $0.9 \times 10^{-4}$ \\
        Si      & $3.2 \times 10^{-5}$ & $\cdots$             & $3.2 \times 10^{-5}$ \\
        \hline
    \end{tabular}
\end{table}

\begin{table}[ht]
    \centering
    \caption{Species included in our model ordered by volatility, with their binding energies $E_{\rm bind}$, desorption prefactors $\nu$, and abundances with respect to H.}
    \label{tab:species}
    \setlength\tabcolsep{2pt}
    \begin{tabular}{rlll}
        \hline\hline
        Species & $E_{\rm bind}$ (K) & $\nu$ ($\mathrm{s^{-1}}$) & Initial abundance \\
        \hline
        Si-grains                & -                        & -                     & $\mathrm{Si/H} = 3.2 \times 10^{-5}$ \\
        C-grains                 & -                        & -                     & $0.39  \times \mathrm{C/H} = 9.0\times10^{-5}$ \\
        \ce{H2O}                   & 6722\tablefootmark{a}    & $1.3\times10^{18}$                & $0.2 \times \mathrm{O/H} = 1.1 \times 10^{-4}$ \\
        \ce{CH3OH}        & 4850\tablefootmark{b}    &$5.0\times10^{14}$     & $0.01  \times \mathrm{C/H} = 2.3\times10^{-6}$ \\
        \ce{CO2}                   & 2980\tablefootmark{c}    & $1.1\times10^{15}$\tablefootmark{e}    & $0.09  \times \mathrm{C/H} = 2.1\times10^{-5}$\\
        \ce{CH4} & 1190\tablefootmark{d}    & $2.5\times10^{14}$    & $0.01  \times \mathrm{C/H} = 2.3\times10^{-6}$ \\
        \ce{O2}   & 1030\tablefootmark{d}    & $1.3\times10^{14}$    & $0.16  \times \mathrm{O/H} = 0.85\times10^{-4}$ \\
        CO                       & 910\tablefootmark{d}     & $4.1\times10^{13}$    & $0.50  \times \mathrm{C/H} = 1.2\times10^{-4}$ \\
        \hline
    \end{tabular}
    \tablebib{
    \textsuperscript{a}\citet{Smith_2011},
    \textsuperscript{b}\citet{Doronin_2015},
    \textsuperscript{c}\citet{Edridge_2013},
    \textsuperscript{d}\citet{Smith_2016}
    }
    \tablefoottext{e}{Assuming $1\times10^{15}\,\mathrm{mol\,cm^{-2}}$}
\end{table}

\subsubsection{Elemental abundances}
The elemental abundances of planet-forming material are somewhat uncertain depending on what they are assumed to follow.

For example, the solar value \citep[most recently estimated at $\mathrm{C/H}=2.9\times10^{-4}$][]{Asplund_2021} is often used but is generally found to be higher than measurements of nearby young stars or the interstellar medium (ISM) suggesting the Sun is somewhat carbon-enriched, perhaps as a result of its formation location or environment, or galactic chemical evolution. For example, measurements of B stars \citep{Przybilla_2008,Nieva_2012}, which are short-lived and therefore representative of contemporary star-formation, show median values of $\mathrm{C/H}=2.1\times10^{-4}$.
Similarly, ISM measurements lie even lower at $\mathrm{C/H}\approx1.4\times10^{-4}$ \citep{Cardelli_1996,Savage_1996,Sofia_2004} though this is understood as being due to some C being locked up in the solid phase.
Nevertheless, recombination lines in the Orion Nebula suggest a higher gas-phase abundance of $\mathrm{C/H}\approx2.3\times10^{-4}$ \citep{Simon-Diaz_2011}, which is suggestive of the return of the refractory components to the gas phase \citep{Nieva_2012} in regions undergoing star formation.
As this is intermediate between the B-star and solar inferences, we adopted this as an intermediate estimate of the total (volatile+refractory) elemental carbon budget. 

However, the Orion Nebula appears depleted by an order of magnitude in silicates relative to solar, comparable to the ISM. Therefore, we cannot make the same assumption that refractory silicon carriers have sublimated as for the refractory carbon material. 
In this case, as cosmic abundances are highly consistent with solar, we simply adopted $\mathrm{Si/H}=3.2\times10^{-5}$ \citep{Nieva_2012,Asplund_2021} for the total Si abundance, all of which is assumed to be refractory.

Finally, the oxygen abundance resulting from the sum of the Orion Nebula value $4.5\times10^{-4}$ and that which is locked up in silicates given the assumed silicate abundance and a 3:1 ratio is $\sim 0.9\times10^{-4}$ is $5.3\times10^{-4}$. This value is intermediate between the present solar value \citep[$4.90\times10^{-4}$][]{Asplund_2021} and both the protosolar value \citep[$5.62\times10^{-4}$][]{Asplund_2021} and the B-star estimate of cosmic abundances \citep[$5.75\times10^{-4}$][]{Nieva_2012}; as it is consistent with both sets of values within the uncertainties, we adopted a total (volatile+refractory) $\mathrm{O/H}=5.3\times10^{-4}$.

Our adopted abundances are summarised in Table\,\ref{tab:elements}.

\subsubsection{Molecular abundances}
\label{sec:elementalBudgets}
Correctly assigning the total elemental budgets to their various carriers is a non-trivial job. Adding up ices and expected volatiles typically falls comfortably short \citep[e.g.][]{Boogert_2015,McClure_2023}; however, when including refractory sources one has to be simultaneously mindful of species that are very hard to constrain without distinct spectral features (such as amorphous carbon), and the risk of double counting species as not all spectral features are unambiguously assigned.

We assumed that the difference between our adopted carbon abundance and the ISM value is representative of the total refractory carbon, which suggests that around 40\% of the carbon is in refractory material of some kind.
For the volatile C-carrying molecules, we base our abundances on ice observations towards protostellar envelopes and molecular clouds \citep{Oberg_2011b,Boogert_2015}. This only totals around 30\% of the expected volatile carbon budget; we assume the missing C resides in CO, and add this to its contribution from the ices. The resulting total CO abundance is close to the total estimated from gas and ices for low-mass protostars \citep{vanDishoeck_2021}.

For oxygen, we adopted an assumption that 20\% of the total oxygen resides in \ce{H2O}.
This is a factor 2-3 lower than expected by models \citep{Bergin_2000,Walsh_2015} which predict highly efficient and rapid formation of water by grain-surface chemistry.
Moreover, it is up to a factor 2-3 higher than derived for ices in protostellar envelopes or molecular clouds \citep{Pontoppidan_2004,Boogert_2015,McClure_2023}.
Dust growth is one plausible contributor to the low values: ices on larger grains, which could account for around 50\% of the water ice, can become invisible to infrared absorption \citep{vanDishoeck_2021}. However, as there is no direct evidence for this water, for the sake of tracking the missing oxygen in the code, we assume that the remaining 32\% of the oxygen resides in \ce{O2}. Even though this is somewhat larger than predicted by thermochemical disc models \citep{Walsh_2015}, and in tension with observed upper limits of a few \% \citep{vanDishoeck_2021}, this does not affect our results as we do not model its chemistry or emission.

\subsubsection{Molecular data}
Our chemical model is thus based on the Case 2 scenario of \citet{Booth_2017}; however, we include some additional species, updated some of the initial abundances and binding energies, and used prefactors in the desorption rates derived from experiments rather than the approximation of \citet{Hasegawa_1993}.
The updated values for each species are listed in Table\,\ref{tab:species}.

For the most abundant ices central to this work such as \ce{H2O} and \ce{CO2}, we expect many monolayers to be present on dust grain and thus we take our binding energies from multilayer Temperature Programmed Desorption experiments (which all correspond to pure ices). Only once the ice depletes to the final monolayer(s) will the binding energies become affected by the substrate and begin to deviate; although this will have little effect for the majority of the ice budget.
For consistency, the binding energies are used in conjunction with prefactors derived simultaneously which is important for correctly identifying the location of the snow line \citep{Minissale_2022}. These data are all taken from works included in Tables 2, 3, and 5 of the review by \citet{Minissale_2022} with full details given in the notes of Table\,\ref{tab:species}.

Thus, an implicit assumption is that the ices are pure and successively layered, each with a single desorption temperature. However observations of protostellar envelopes find that several components are needed to fit ice absorption bands, implying that the molecules may exist in several ice phases \citep[see review by][]{Boogert_2015}. In particular, for \ce{CO2}, up to five components are needed \citep{Pontoppidan_2008,Brunken_2024}, of which the 
pure phase is never seen to constitute more than about 20\% (and is more typically <10\%), while the polar phase (mixed with \ce{H2O}) is overall the most abundant \citep{Oberg_2011b,Brunken_2024}.
If all \ce{CO2} resided in the polar phase, it could become trapped below frozen material even above its pure-phase desorption temperature. However, such \ce{CO2} does not only codesorb at the \ce{H2O} desorption temperature, but much of it diffuses to the surface and sublimates; the rest is released at a temperature in between the \ce{CO2} and \ce{H2O} sublimation temperatures in a process of `volcano desorption' taking place during the amorphous-to-crystalline \ce{H2O} phase transition \citep{Collings_2004,Edridge_2013}.
Moreover, thermal processing can lead to clusters of pure \ce{CO2} forming from the polar phase in a process known as segregation \citep{Ehrenfreund_1999,Oberg_2009c}, shifting the balance towards pure ices on longer timescales.
Thus, given the likely mix of ice phases, possible segregation from the polar phase, and the release of \ce{CO2} below the \ce{H2O} sublimation temperature even in the polar phase, we nevertheless expect a separation between the locations where \ce{H2O} and \ce{CO2} sublimate in the disc; this work assumes the most extreme scenario. 

\FloatBarrier

\section{Synthetic spectrum model}
\label{sec:methodsSpectrum}
\subsection{The origin of MIR emission}
For many molecules, rotational, vibrational, and ro-vibrational transitions produce line emission in the mid-infrared. For \ce{CO2}, the strongest feature - often the only one clearly detected - is the Q-branch at $15\,\mathrm{\mu m}$ corresponding to purely vibrational excitation of the bending mode. In more \ce{CO2}-dominated spectra, it can be flanked by the R- and P-branches from ro-vibrational excitation of the bending modes and the `hot bands' at $13.85\,\mathrm{\mu m}$ and $16.2\,\mathrm{\mu m}$ - the Q-branches for transitions between bending and stretching modes. \ce{H2O} shows purely rotational transitions at wavelengths $\gtrsim 10\,\mathrm{\mu m}$, and ro-vibrational transitions with higher excitation energies (thus tracing hotter gas) at shorter wavelengths. These lines have typical excitation temperatures $\gtrsim 1000\,\mathrm{K}$ and so trace regions with temperatures of 100s K. Thus, the most strongly emitting regions would be in the vicinity of the disc's optical photosphere, where the temperatures rise, and/or the inner $\lesssim 1\,\mathrm{au}$. This can be seen in 2D thermochemical models \citep{Woitke_2018,Bosman_2022a,Bosman_2022b}, where the majority of line emission is from a thin hot layer. This layer is exposed to UV and undergoes rapid photochemical processing, and is not necessarily representative of the bulk composition towards the midplane \citep{Anderson_2021} that we follow with our 1D models. However, vertical transport and mixing may still lead to a signature in the surface of any midplane enhancement, depending on how the gas is reprocessed.

As well as the simple case of excitation determining the dominant emitting layer, there are three further features that limit the depth to which emission can be seen. Firstly, the dust in the disc provides a continuum opacity, which hides material beyond the MIR photosphere. Moreover, if there is a drop in the abundance vertically, for example when crossing the snow surface, emission may be potentially visible down to that depth but no further due to a lack of molecules to emit. Finally, the emission of strongest transitions may become optically thick due to the line opacity after a relatively small column density of material and no longer grow with an increased molecular column density; in this case changes in dust continuum opacity have less effect \citep{Bruderer_2015}.
This cooler material between the hot surface and the photosphere or snow surface may, in principle, be traced with an appropriate choice of lines, by picking those with a lower excitation energy which may be penalised by the partition function at high temperatures, or those which have low Einstein coefficients and thus can avoid becoming optically thick in the hot surface layer \citep[e.g.][]{Gasman_2023}. However in practice, optically thin lines may be rare \citep[][who found four clean, non-blended lines out of a total of >100, all tracing the hottest gas]{Temmink_2024b}.

Moving beyond individual lines, the most widely used tools for the interpretation of MIR spectra are 0D LTE slab models. These are characterised by a column density, temperature, and emitting radius (itself a parametrization of the emitting area which simply scales the spectrum to match the observed fluxes) \citep[e.g.][]{Carr_2011,Salyk_2011}.
Although many works start by fitting a single slab model to the emission from each molecule, 2-3 components are typically needed to fully represent the \ce{H2O} rotational spectrum, fitted either simultaneously using multiple slabs \citep{Temmink_2024b}, to different parts of the spectrum \citep[e.g.][]{Gasman_2023}, or to the residuals after another disc spectrum is subtracted as a template \citep{Banzatti_2023}; this allows weaker emission components to be recovered.

Typically, the dominant rotational \ce{H2O} component ---and the one which a single slab finds--- is the hottest, with $T \approx 600-800\,\mathrm{K}$. This must represent the hot disc surface described and indeed matches the temperatures predicted for this layer by \citet{Bosman_2022a} when chemical heating is accounted for.
The coldest component, present only in some discs, is found at $\lesssim 200\,\mathrm{K}$ and is typified by prominent emission lines at $23.817$ and $23.896$ \citep{Banzatti_2023}. Given its proximity to the desorption temperature of water, it may represent sublimating ices and has thus been interpreted as a signal of drift. However, emission purely in the vicinity of the midplane snow line would not reproduce the large emitting radius inferred \citep{Banzatti_2023,Temmink_2024b} when these lines are prominent. 
Thus, this component must trace the full 2D nature of the snow surface, something that cannot easily be captured in our 1D models which assume a vertically isothermal disc. 
Between these two components is an intermediate `warm' component, typically found at $250-500\,\mathrm{K}$. This is the component that our 1D models are likely to mostly closely match in the absence of the full 2D thermochemical structure needed to understand the other components (see also Appendix \ref{appendix:Ch4}). However in practice, all its clean emission lines (i.e. detectable above noise, without blending with other features) may still be considerably optically thick \citep[$\tau\sim350$,][]{Temmink_2024b} and hence not trace the full column density to the continuum MIR photosphere.

\subsection{Constructing the spectrum}
We follow an approach similar to \citet{Anderson_2021} to construct a forward model of the MIR spectrum of our molecules of interest - \ce{H2O} and \ce{CO2} - by dividing the disc into an annulus at each modelled radius $R_i$ and summing the individual spectrum of each annulus $F_{\lambda, i}$, weighted by the area of the annulus:
\begin{equation}
    4 \pi d^2 F_\lambda = \sum_i 2 \pi R_i \Delta R_i F_{\lambda, i} 
    ,
\end{equation}
where $d=140\,\mathrm{pc}$ is a typical distance to a nearby disc, and where we sum over all annuli with a non-negligible gaseous reservoir of the element (here taken as all grid cells where $f_{\rm spec, gas}/f_{\rm spec, ice}>10^{-5}$).
We modelled the spectrum over the wavelength range of MIRI-MRS Channels 3B ($13.34-15.57\,\mathrm{\mu m}$) \& 3C ($15.41-17.98\,\mathrm{\mu m}$) which cover the main \ce{CO2} features and the warm water emission 
(see Appendix \ref{appendix:Ch4} for longer wavelength \ce{H2O} emission).

The spectrum of each annulus was given by a 0D LTE slab model calculated using \textsc{radexpy} \citep{Tabone_2023}. As discussed, these are characterised by a temperature $T_i$ and a visible column density of material $N_{\rm vis, i}$. The slab models naturally account for the line opacity, so to model the emission of the bulk material, we must provide properties corresponding to the MIR $\tau=1$ surface which is the approximate maximum depth to which we can trace. Consequently, we took the midplane temperature $T_i=T(R_i)$ given by Eq.\,\ref{eq:Tdisc} assuming the deep layers are approximately vertically isothermal.
Assuming that gas - including the emitting molecules of interest - is vertically well mixed, then the molecules follow a Gaussian distribution in the vertical direction, and the visible column density is given by
\begin{equation}
    N_{{\rm vis},i} = \frac{N_{{\rm tot},i}}{\sqrt{\pi}} \int_{-\infty}^{\infty}{e^{-\tau_i(\tilde{z})} e^{-\tilde{z}^2} {\rm d}\tilde{z}}
    \label{eq:Nvis_full}
    ,
\end{equation}
where $\tilde{z} = z/\sqrt{2H_i}$, $\tau_i(\tilde{z}=\infty)=0$, $\tau_i(\tilde{z}=-\infty)=\tau_i$, and $N_{{\rm tot},i}$ is determined directly from the surface density of each molecule and its molecular mass.
The main source of opacity is the small dust, which is also likely to be well-mixed (while the large grains largely settle to the midplane where they cannot contribute), and at each radius is assumed to have an abundance given by $f_{\rm small}\epsilon$ where $f_{\rm small}=0.03-0.25$ comes from the two-population model, and the total dust-to-gas ratio $\epsilon$ varies spatially and temporally following the dust evolution that we model. In the absence of radial drift, this would imply a dust-to-gas ratio in the upper layers of $3\times10^{-4} - 2.5\times10^{-3}$, but we emphasise this may be raised by the delivery of dust or lowered if it drifts.

Given that the small dust is well-mixed and roughly follows the same vertical profile as the gas, we can understand Eq.\,\ref{eq:Nvis_full} with the approximation that:
\begin{equation}
    N_{{\rm vis},i} \approx N_{{\rm tot},i} \frac{1-e^{-\tau_i}}{\tau_i}
    \label{eq:Nvis_approx}
    .
\end{equation}
In the limit $\tau_i \gg 1$, the visible column density is simply the fraction of the column lying above the MIR $\tau=1$ surface $N_{{\rm vis},i} \approx N_{{\rm tot},i}/\tau_i$.

We note, however, that for full accuracy in our calculation of $\tau(z)$, we did include both the large and small grains, and calculate their vertical profiles include settling according to \citet{Dubrulle_1995}.
For the opacity of the dust we use the opacities from \textsc{dali} \citep{Bruderer_2013} which represent a graphite-silicate mix as per the ISM composition of \citet{Weingartner_2001b}.

The most major limitation is that we assume a vertically constant temperature. In reality, different molecules are seen to present quite different emitting temperatures, with colder temperatures thought to trace lower heights into the disc. In general, \ce{CO2} is expected to trace a deeper layer (i.e. higher column) than \ce{H2O} \citep{Woitke_2018}, and \ce{^13CO2} even deeper (and colder) due to excitation and line optical depth effects. Moreover, in equilibrium thermochemical models, \ce{H2O} is formed at a higher temperature than \ce{CO2} which favours a high \ce{H2O}/\ce{CO2} at high temperatures. A vertical temperature gradient would overall limit the column of material that dominates the emission, especially for the \ce{H2O} that traces the hotter material, and may alter the \ce{CO2}/\ce{H2O} ratio. 
In general, however, the height of optical photosphere which sets the depth of the hotter gas is also controlled by the dust column density, so the general relationship of Eq.\,\ref{eq:Nvis_approx} still holds, just with a higher optical depth.

For \ce{CO2}, we accounted for mutual shielding of lines \citep{Tabone_2023} - known as line overlap - but this is not required for the purely rotational lines of \ce{H2O} \citep{Temmink_2024b}.

\subsection{Interpreting the spectrum}
To interpret our predicted spectra, we compared them to a grid of slab models covering $T=25-1500,\mathrm{K}$ in steps of $25\,\mathrm{K}$ and $\log_{10}(N/\mathrm{cm^{-2}})=16-25.92$ in steps of 0.16 \citep[following][]{Gasman_2023}.
As the flux of a slab model simply scales with the emitting area $A= \pi R_{\rm eff}^2$, for each pair $(T,N)$, the corresponding effective emitting radius $R_{\rm eff}$ is that which minimises the squared difference between the full spectrum and the individual slab model. The best fit $(T,N,R_{\rm eff})$ is that which produces the smallest squared difference across all of the slab models.

As aforementioned, it is now common to fit real spectra with slab models in multiple wavelength bands \citep{Gasman_2023,Schwarz_2024}, or with a 2- or 3-component slab model \citep{Temmink_2024b}, in order to extract information about multiple components. However, as we are anyway only modelling the contribution of warm \ce{H2O} and \ce{CO2}, we only fit a single slab, to our entire modelled wavelength range.
In Appendix \ref{appendix:Ch4}, we test instead using the MIRI-MRS Channel 4 wavelength range ($17.70-27.90\,\mathrm{\mu m}$) for \ce{H2O} but did not find significantly different results; this illustrates how our spectra lack the full complexity of those observed \citep[e.g.][]{Temmink_2024b}.

\FloatBarrier

\begin{figure*}[th]
    \centering
    \includegraphics[width=0.9\linewidth]{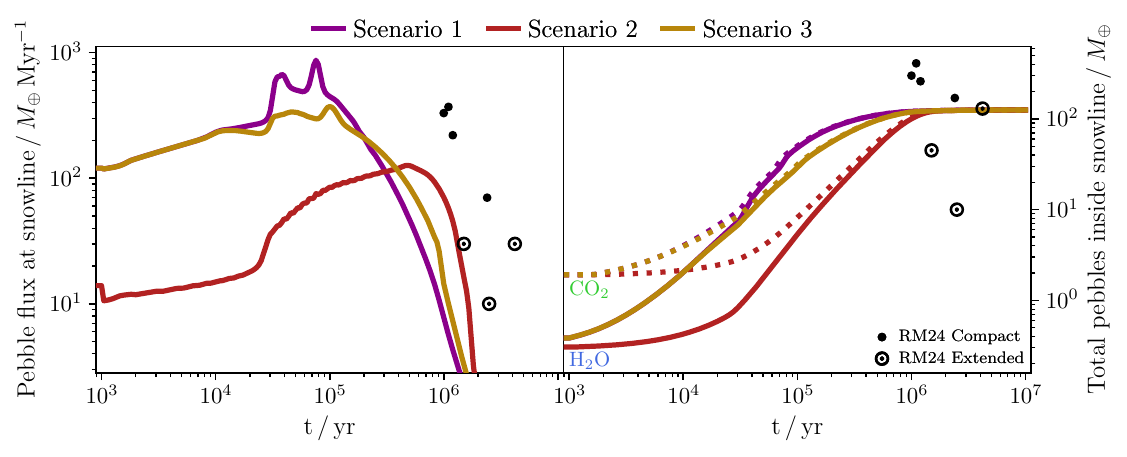}
    \caption{Evolution of the pebble component of the disc. Right: Total dust mass (including that which has crossed the inner disc boundary and accreted onto the star) inside the initial locations of the \ce{H2O} (solid) and \ce{CO2} (dashed) snow lines (right). Left: Pebble flux across the \ce{H2O} snow line. Included for reference are the values from Table 6 of \citet{RomeroMirza_2024b} [RM24], which are estimated from the total number of \ce{H2O} molecules $<400$\,K, split into compact discs (small, filled, markers) and extended discs (larger, ringed, markers).}
    \label{fig:pebbleFlux}
\end{figure*}

\begin{figure*}[th]
    \centering
    \includegraphics[width=0.9\linewidth]{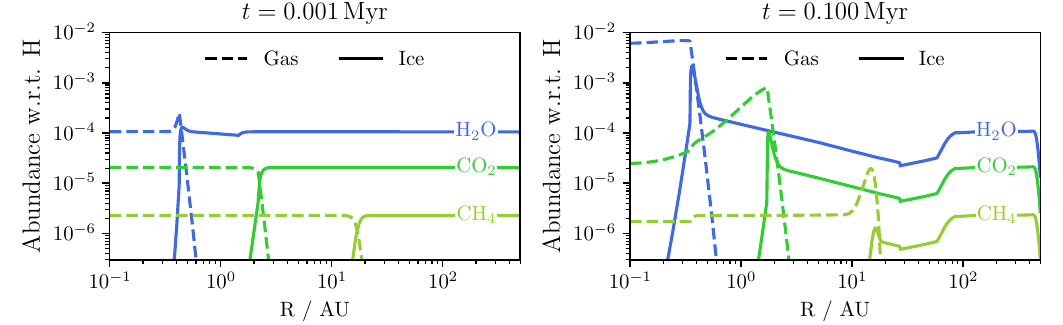}
    \caption{Radial distribution of the \ce{H2O}{} (blue), \ce{CO2}{} (bright green), and \ce{CH4}{} (olive green) after 1 kyr of evolution (left) and 0.1 Myr of evolution in a smooth disc in Scenario 1. Solid lines indicate molecules in the solid (ice) phase, and dashed lines molecules in the gas (vapour) phase.}
    \label{fig:snowlines}
\end{figure*}

\section{Results}
\label{sec:results}
\subsection{Evolution of smooth discs}
We begin by briefly showing the evolution of smooth discs.
To contextualise this discussion, Fig. \ref{fig:pebbleFlux} shows the evolution of the cumulative dust mass that has reached inside of \ce{H2O} and \ce{CO2} snow lines on the right-hand panel and its rate of change, i.e. the pebble flux across the \ce{H2O} snow line, on the left-hand panel. As the \ce{CO2} snow line is further out in the disc, there are initially more solids inside it. The pebble flux peaks at around 0.1\,Myr in Scenarios 1 \& 3, but somewhat later in Scenario 2. As we preserve the degree of coupling ($S\!t/\alpha$) between Scenarios 1 \& 2, the Stokes number is smaller in the latter case, and the pebbles are delivered more slowly, over a longer period.

\subsubsection{Spatial and temporal evolution of molecules}

Figure \ref{fig:snowlines} shows the radial profile in the ice and gas phase of three major molecules - \ce{H2O}, \ce{CO2} and \ce{CH4}{} - at two different times with Scenario 1 for the gas and dust transport. We see that water, being a polar molecule with a high binding energy, is only in the gas phase within around $0.47\,\mathrm{au}$ of the star. \ce{CO2} is next to freeze-out, at $\sim2.2\,\mathrm{au}$. Finally, \ce{CH4} is the most volatile and is only frozen out outside of $16\,\mathrm{au}$. After only 1000 years (left panel), there has been little redistribution of the molecules. After 0.1 Myr, when the pebble flux peaks, drift is possible for grains out to around 80 au i.e. across essentially the whole disc, resulting in a reduction in the ice abundances represented by the solid lines. For example, an amount of \ce{CH4} ice has been moved from 16-80 au to the snow line, but it has not had enough time to spread further inwards and is thus concentrated in a small, narrow peak there. In contrast, \ce{CO2} has been moved from a slightly wider area (2-80 au) leading to a larger peak, and the local viscous timescale is shorter at its snow line than for \ce{CH4}, and so it has spread further in. \ce{H2O}{} has been depleted from the largest area and so shows the greatest enhancement, and the peak has already spread so much as to produce a uniform abundance inside the snow line.

\begin{figure}[th]
    \centering
    \includegraphics[width=0.9\linewidth]{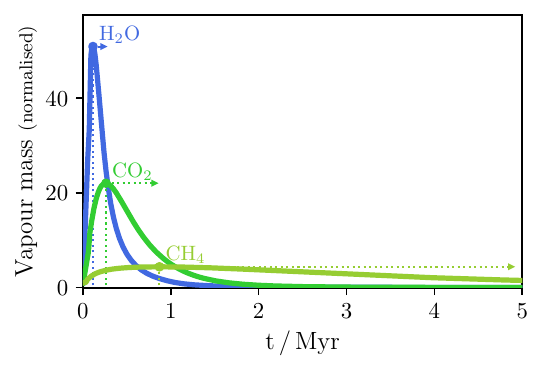}
    \caption{Evolution of the vapour mass of \ce{H2O}{} (blue), \ce{CO2}{} (bright green), and \ce{CH4}{} (olive green) normalised to their initial values for a smooth disc in Scenario 1. The vertical dotted lines mark the times that each molecule reaches its peak vapour mass and the rightward arrows indicate the timescale for the peak to then decrease (Eq. \ref{eq:taccrete})}
    \label{fig:vapourEvolution}
\end{figure}

Figure \ref{fig:vapourEvolution} translates such profiles into the total vapour mass of different molecules as a function of time. Each sees an enhancement as it is delivered to its snow line by radial drift, followed by a decline once drift delivery slows and the deposited vapour is advected onto the central star.
The molecules undergo this enhancement in order of their volatility. This happens because the less volatile the molecule, the smaller its initial vapour reservoir but the larger its initial ice reservoir and thus it can undergo a larger and more rapid enhancement. Moreover, pebbles start to drift at $R\gtrsim 0.47\,\mathrm{au}$ and deliver \ce{H2O}{} to its snow line after a fairly short growth timescale, whereas dust at larger radii takes longer to grow to drifting sizes \citep[][Eq. 13]{Birnstiel_2012}, and so delivery to more distant snow lines starts after a short delay.

Finally, the decline from the peak enhancement is driven by the accretion of each molecule onto the star: the timescale for this to happens for molecule X can be approximated as
\begin{align}
    t_{\rm accrete} &:= M_{\rm vap, X}/\dot{M}_{\rm acc, X} \\
                    &= \frac{2\pi \Sigma_{\rm snow} R_{\rm snow}^2 A_{\rm X}}{3\pi \Sigma_{\rm snow} \nu_{\rm snow} A_{\rm X} } \\
                    &= \frac{2}{3}\frac{R_{\rm snow}^2}{\nu_{\rm snow}} \\
                    &= 0.25\,\mathrm{Myr}\, R_{\rm snow} \left(\alpha/10^{-3}\right)^{-1}
                    ,
    \label{eq:taccrete}
\end{align}
for our choice of stellar mass and disc temperature profile, where $A_{\rm X}$ is the (assumed constant due to diffusion) abundance of molecule X in the inner disc, $R_{\rm snow}$ is the radius of the snow line in AU and $\Sigma_{\rm snow}$ and $\nu_{\rm snow}$ are the gas surface density and viscosity respectively, both evaluated at the snow line. This is of the order of the viscous timescale at the molecule's snow line.
Since the viscous timescale for advection increases linearly with radius, then the vapour of a less volatile element with a closer snow line is drained onto the star more quickly, thus also ending the period of enhancement sooner. 

We thus see a sequence where the disc starts off \ce{H2O}-dominated, evolves through a \ce{CO2}-dominated phase, and finally ends dominated by the most volatile molecules such as \ce{CH4} and CO. According to this sequence, the C/O ratio will also broadly increase over the disc's evolution. Moreover, the timescale to go through this sequence is shorter for lower-mass stars which host colder discs with closer-in snow lines \citep{Mah_2023}. Equation \ref{eq:taccrete} also predicts that the time by which \ce{H2O} is removed by advection and the disc switches to the \ce{CO2}-rich phase will be later - and the subsequent lifetime of the \ce{CO2}-rich phase will be longer - at lower viscosities.

\subsubsection{Modelled spectrum and fits}
\begin{figure*}[htp]
    \centering
    \includegraphics[width=0.9\linewidth]{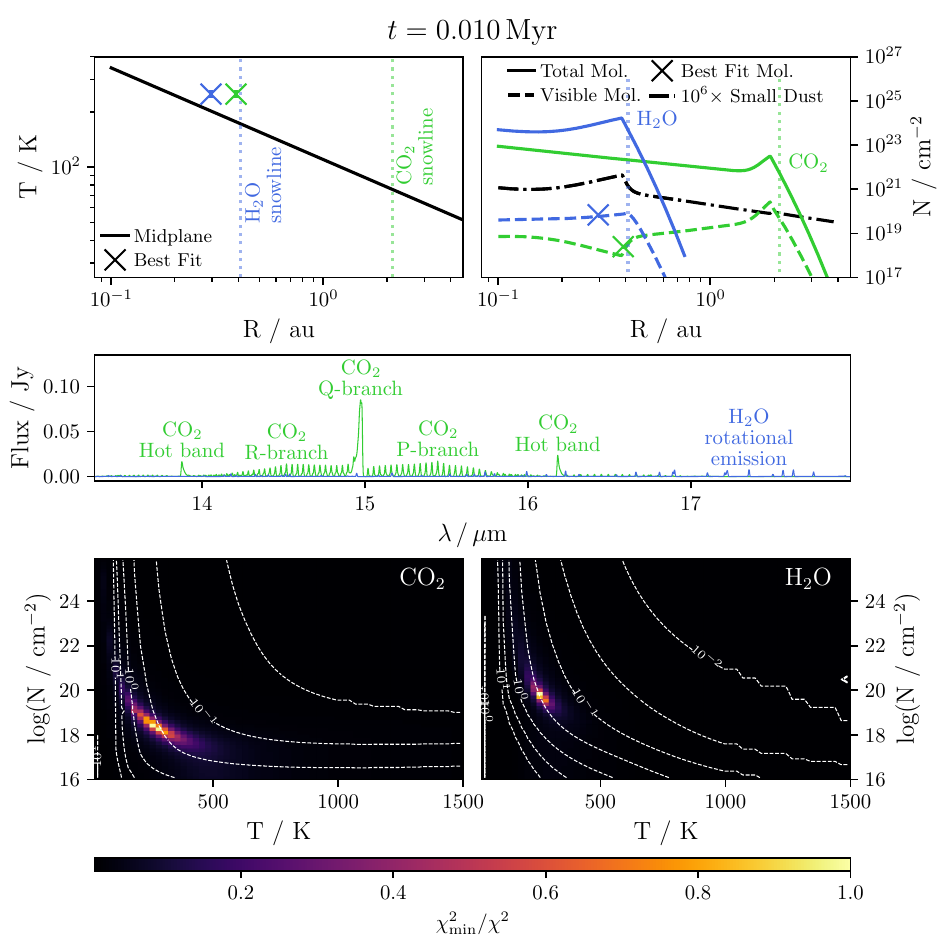}
    \caption{Example of the construction of a synthetic spectrum and the retrieval of parameters using 0D LTE slab model fits. The top panels show the temperature profile (left) and the column density profiles after $10^4\,\mathrm{yr}$ (right) for small dust (black) and total (solid) and visible (dashed) \ce{H2O} (blue) and \ce{CO2} (green). The light, vertical, dotted lines indicate the snow lines. The middle panel shows the synthetic spectra of each molecule (assuming a distance of 140 pc) while the bottom panels show the goodness of fit for a grid of slab models with varying temperature and column density. The white contours show the radius (in AU) of a circle whose area maximises the goodness of fit for each combination of temperature and column density. The overall best fit parameters are shown on the top panels as the crosses, with the vertical error bars representing the spacing of the grid of slab models.}
    \label{fig:exampleSpectrum}
\end{figure*}

In Fig.\,\ref{fig:exampleSpectrum} we show an example of the method we outlined in Sect. \ref{sec:methodsSpectrum}, applied at an age of $10^4\,\mathrm{yr}$ to the same Scenario 1 disc. Appendix \ref{appendix:later} discusses examples at later times of 0.1 Myr \& 1 Myr, shown in Figs. \ref{fig:exampleSpectrumLate} \& \ref{fig:exampleSpectrumLater} respectively.

\begin{figure*}[th]
    \centering
    \includegraphics[width=0.9\linewidth]{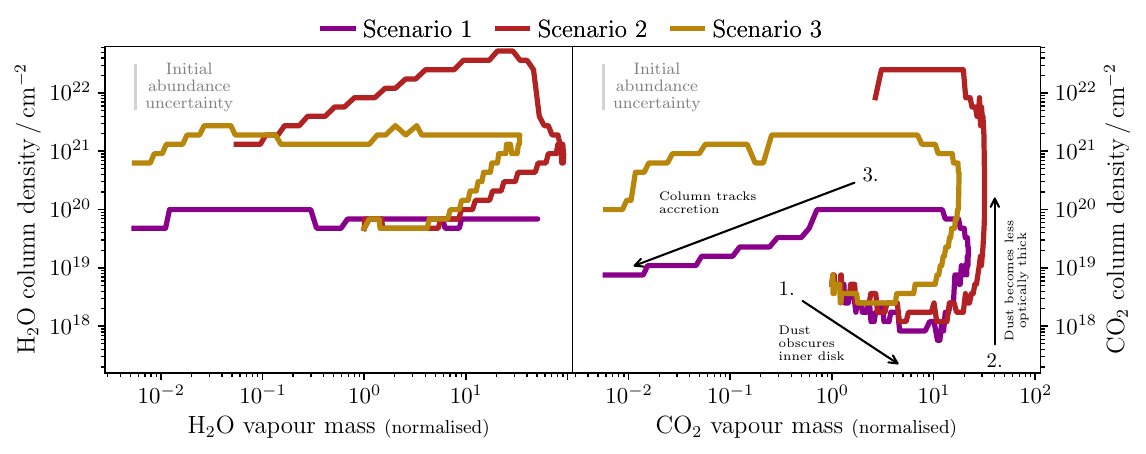}
    \caption{Relationships between the vapour mass and the column density measured using slab fits for \ce{H2O} (left) and \ce{CO2} (right) for a smooth disc model in Scenarios 1 (purple), 2 (gold), and 3 (red). The curves shown are traversed anticlockwise.}
    \label{fig:calibration}
\end{figure*}

In the top right panel, we show that the visible column densities of \ce{H2O}{} and \ce{CO2}{} (dashed lines) as a function of radius are much lower than the total column densities (solid lines), in this case by around four orders of magnitude, as the inner disc is very optically thick and most of the molecular reservoir is therefore hidden. Moreover, while the total \ce{H2O}{} column density has begun to show a peak inside of the \ce{H2O}{} snow line (vertical blue line), reflecting the action of drift delivery as seen in the abundances in Fig.\,\ref{fig:snowlines}, this feature is not apparent in the visible \ce{H2O}{} column density, as the dust that delivered it contributes to increasing the local optical depth. This occurs in this scenario because the dust fragments, encounters a traffic jam\footnote{These traffic jams are distinct from the \citet{Mah_2024} `traffic jam scenario', which describes the leakiness of traps of intermediate depth.}, and thus moves at the same rate as the liberated vapour, resulting in the cancellation of the effects of enhancement and increased optical depth. The increased dust obscuration is also seen in the visible \ce{CO2}{} column density, which displays a dip inside of the water snow line. Conversely, the increase in \ce{CO2}{} inside its snow line is reflected in the visible column density as there is no correlated local pile up of dust.

The central panel shows channels 3B and 3C of the MIRI-MRS spectra we construct for each molecule from the temperature profile (top left) and visible column densities (top right). This particular case presents here as a \ce{CO2}-dominated spectrum due to its lack of a modelled hot water component and its fairly large \ce{CO2} emitting area and relatively cool water emitting temperature compared to observations.
The bottom panels show the $\chi^2$ maps for the slab fits, with brighter colours indicating a better fit.
We then overplot the parameters of the single best fit amongst these models --- $T,N,$ and $R_{\rm eff}$ --- on the temperature profile and the column density profiles. For \ce{H2O}, the fit seems to trace a radius a little inside of the snow line, suggesting that it is a reasonably good measure of the area over which \ce{H2O} is in the gas phase. For \ce{CO2} however, the radius is very similar, suggesting that \ce{CO2} near the snow line is a bit too cold to contribute much to the spectrum. However, we do see the radius increase somewhat later in the evolution as the pile up of dust due to the traffic jam increases and the contrast between the visible \ce{CO2} column density inside and outside the water snow line grows such that the cold \ce{CO2} contributes relatively more to the spectrum (see e.g. Fig. \ref{fig:exampleSpectrumLate}). For \ce{H2O}, we observe that the retrieved column density agrees well with the input visible one. For \ce{CO2}, the visible column densities are more radially variable, and the column density ends up as some sort of average. Similarly, in both cases, the retrieved temperature is somewhat higher than $T(R_{\rm eff})$ as it is biased by the warmer gas in the inner disc. Consequently for both molecules the temperatures consistently lie in the $200-300\,\mathrm{K}$ range consistent with an area-weighted average $\langle T \rangle = \frac{4}{3} \frac{R_{\rm eff}^{3/2}-R_{\rm in}^{3/2}}{R_{\rm eff}^{2}-R_{\rm in}^{2}} T_{\rm in} \approx235\,\mathrm{K}$ given $R_{\rm in}=0.1\,\mathrm{au}$, $T_{\rm in}=T(R_{\rm in})=347\,\mathrm{K}$, and $R_{\rm eff}\approx0.32\,\mathrm{au}$.

\subsubsection{Sensitivity of column density to underlying abundance}
Figure \ref{fig:calibration} shows how the observed column densities of \ce{H2O} and \ce{CO2} depend on their vapour masses over the course of the disc evolution, now for smooth discs in all three scenarios.

In Scenario 1, the measured water column density is almost entirely insensitive to the true vapour mass and underlying column density. As discussed above (in the context of the lack of a clear delivery signature in the visible water column density at $10^4\,\mathrm{yr}$), this occurs because both the water vapour released by the dust grains that cross the snow line - and the dust grains themselves - move with the gas inside of the snow line. Therefore, while radial drift enhances the mass of water in the inner disc, it also enhances the amount of dust and makes the disc more optically thick. The water vapour is thus hidden by the very same dust that delivers it.
This constant value of the visible water column density can be explained quite simply. When the disc is highly optically thick, given that the opacity is dominated by the well-mixed small dust grains, then (see Sect. \ref{sec:methodsSpectrum}):
\begin{align}
    N_{\rm H_2O, vis} &\approx \frac{N_{\rm H_2O, tot}}{\tau} \\
                &\approx \frac{\Sigma_{\rm H_2O, tot} / (m_{\rm H_2O})}{f_{\rm small} \Sigma_{\rm dust} \kappa_{\rm small}} \\
                &\approx 7.2\times10^{19} \frac{\Sigma_{\rm H_2O}/\Sigma_{\rm dust}}{0.54}  \left(\frac{\kappa_{\rm small}}{1000\,\mathrm{cm^{2}\,g^{-1}}}\right)^{-1}
                \label{eq:expectedNvis}
                .
\end{align}
Since the small grains do not evolve in size, then their opacity is fixed. Moreover, since there is a traffic jam, then dust and water vapour are transported at the same rate and hence $\Sigma_{\rm H_2O}/\Sigma_{\rm dust}$ is the same inside the snow line as when the water was in the ice phase outside the snow line. Hence, this ratio is mainly set by the initial water ice abundance on the dust grains, which we assumed to be a constant throughout the disc. As discussed in Sect. \ref{sec:elementalBudgets}, observations of ices in protostellar envelopes and molecular clouds typically imply a value 2-3 times lower, while models predict one 2-3 times higher if all oxygen goes into \ce{H2O} ice. Therefore, for example, if the discs uniformly inherit the typical protostellar values, then the observed column would be lower by this factor. Nevertheless, this value can serve as a benchmark. If the column densities are much higher, it implies either a loss of inner disc dust with respect to water, or an increase in the water due to processes unrelated to dust delivery.

One way to lose inner disc dust with respect to water is if it does not couple to the gas but continues to undergo radial drift.
Figure \ref{fig:calibration} shows that in Scenarios 2 and 3, where there is no change in the dust properties across the snow line and hence no traffic jam (so $\Sigma_{\rm H_2O}/\Sigma_{\rm dust}$ is no longer constant) the visible \ce{H2O} column density increases by over an order of magnitude. There are times where column density and vapour mass increase or decrease together, but during periods where the loss of accreted vapour is compensated by a loss of drifting dust (thus exposing more \ce{H2O} vapour) the result can be further regions of insensitivity to water vapour mass (Scenario 3) or even regions where vapour mass and column density anti-correlate (Scenario 2).

While \citet{Bosman_2017} argued that the \ce{CO2} emission is degenerate to changes in the global dust-to-gas ratio or \ce{CO2} abundance, here \ce{CO2} shows a cleaner dependence of its retrieved column density on its underlying vapour mass in all cases, though again there is no simple one-to-one relationship. This is because of the different dust dynamics - and hence evolution of the dust-to-gas ratio - at radii inside and outside the \ce{H2O} snow line. Initially, although the vapour mass starts to rise, mainly just inside the \ce{CO2} snow line, the delivery of dust obscures the warmer and brighter \ce{CO2} inside the \ce{H2O} snow line and the visible column there reduces a little, as seen in the above example of the spectrum construction at 0.01 Myr. As the obscured material is hotter than the delivered material, it contributes more to the line fluxes and thus the retrieved column densities are decreased overall. This effect is strongest in Scenario 1 where the traffic jam causes the dust to pile up more. 
However, once supply of drifting dust drops, the dust remaining between the \ce{H2O} and \ce{CO2} snow lines drifts inwards, making this region less optically thick until it dominates the emission. Hence, the retrieved column density rises by at least two orders of magnitude, doing so more rapidly the greater the difference between the dust drift timescale and the gas accretion timescale. Finally, as the \ce{CO2} is eventually removed by advection onto the star, the retrieved column density also drops.

We conclude that \ce{CO2} is a somewhat more promising tracer of the enhancement by drift on timescales of $\gtrsim1\,\mathrm{Myr}$ since it is not so susceptible to observability effects caused by the potential coupling of the dust to the gas inside the \ce{H2O}{} snow line. However, the parameters of the dust and gas evolution still introduce a lot of uncertainty in the relationship between an observed column density and the underlying vapour mass.

\subsection{Evolution of gapped discs}

\begin{table*}[t]
    \centering
    \caption{Discs around stars $>0.3\,M_{\sun}$ observed with MIRI-MRS to which we compare our models.}
    \label{tab:observed}
    \begin{tabular}{rlllll}
        \hline\hline
         Disc   & $M_*\,/\,M_{\sun}$    & Age\tablefootmark{a} / Myr & $N_{\rm H_2O}\,/\,\mathrm{cm^{-2}}$   & $N_{\rm CO_2}\tablefootmark{b}\,/\,\mathrm{cm^{-2}}$   & Reference   \\
        \hline
         GW Lup & 0.46                  & $2\pm1$   & $3.2\times10^{18}$                    & $2.2\times10^{18}$    & \citet{Grant_2023}    \\
                &                       &           &                                       & $6.8\times10^{18}$\tablefootmark{c}    & \\
         Sz 98  & 0.74                  & $2\pm1$   & $3.7\times10^{18}-7.5\times10^{19}$   & $2.4\times10^{18}$    & \citet{Gasman_2023}   \\
         SY Cha & 0.70                  & $1.5\pm0.5$ & $7.0\times10^{18}-3.1\times10^{21}$   & $4.9\times10^{17}$  & \citet{Schwarz_2024}   \\
         AS 209 & 0.96                  & $1\pm0.5$ & $4.7\times10^{17}-2.9\times10^{19}$   & $<5.5\times10^{16}$   & \citet{RomeroMirza_2024}   \\
         CX Tau & 0.37                  & $2\pm1$   & $4\times10^{19}$                     & $7\times10^{17}$    & \citet{Vlasblom_2024b}    \\
                &                       &           &                                       & $1\times10^{19}$\tablefootmark{c}    & \\
        DR Tau & 0.93                   & $2\pm1$   & $4.0\times10^{17}-1.6\times10^{19}$\tablefootmark{d}   & $2.5\times10^{17}$  & \citet{Temmink_2024a,Temmink_2024b}   \\
        DoAr 33 & 1.1                   & $4\pm2$   & $3.2\times10^{18}$                    & $6.3\times10^{17}$    & \citet{Colmenares_2024}    \\
         \hline
    \end{tabular}
    \newline
    \tablefoottext{a}{Ages are mostly indicative of those typically assumed for the star forming regions where the discs are located.}
    \tablefoottext{b}{Where a range of column densities is indicated, it corresponds to the values obtained by fitting to different sections of the spectrum or multiple components. For the other discs, one can expect up to an order of magnitude uncertainty in either direction depending on fit windows chosen and the strength of the emission.}
    \tablefoottext{c}{Alternative estimate from \ce{^13CO2} assuming an isotope ratio of 68.}
    \tablefoottext{d}{Multiple-component fit, Approach III}
\end{table*}

\subsubsection{Scenario 1: Discs with traffic jams}
\begin{figure*}[ht]
    \centering
    \includegraphics[width=0.9\linewidth]{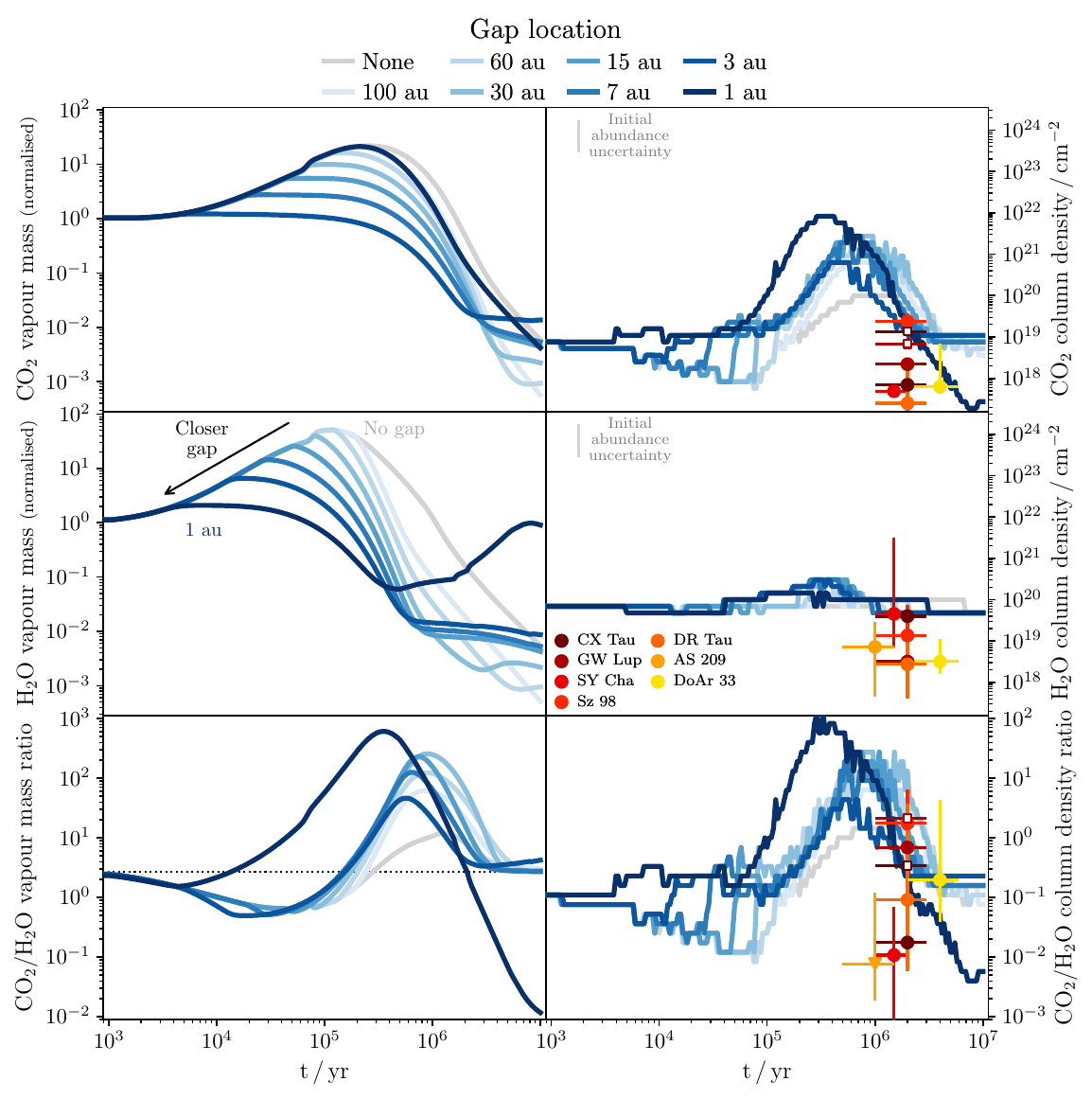}
    \caption{Evolution of the vapour mass (left) and column density (right) for \ce{CO2} (top row), \ce{H2O} (middle row), and their ratio (bottom row) for discs with traffic jams inside the water snow line (Scenario 1). The grey line represents a smooth disc with no traps, while the blues lines, in order of increasingly dark shade, represent traps with increasingly close in traps. Overplotted at their approximate age are the results of slab model fits for a few discs from the literature: CX Tau \citep{Vlasblom_2024b}, GW Lup \citep{Grant_2023}, SY Cha \citep{Schwarz_2024}, Sz 98 \citep{Gasman_2023}, DR Tau \citep{Temmink_2024a,Temmink_2024b}, AS 209 \citep{RomeroMirza_2024} and DoAr 33 \citep{Colmenares_2024}, ordered by increasing stellar mass (more yellow colours). The open squares indicate an alternative estimate obtained from the \ce{^13CO2} isotopologue. The grey bar in the top left shows the $\pm\log_{10}(2.5)$ uncertainty depending on the basis used for the initial ice abundances (Sect. \ref{sec:elementalBudgets}).}
    \label{fig:scenario1evolution}
\end{figure*}

We now present the full results for our first parameter set, which leads to the traffic jam inside the snow line. Figure \ref{fig:scenario1evolution} shows the evolution of the vapour masses and retrieved column densities for models with different gap radii.
As discussed, in the smooth disc, the \ce{H2O} vapour mass rises more quickly than that of the \ce{CO2}. This is reflected in an initial drop in the \ce{CO2}/\ce{H2O} vapour mass ratio, followed by a subsequent increase as the \ce{H2O} vapour mass starts to decline. At the latest times, the models tend back towards the initial ratio again as then the main source of both molecules is sublimation from remnant small grains that are fully entrained with the gas and so do not deliver \ce{H2O} to the inner disc any faster than the gas delivers \ce{CO2}.

Including a gap in the disc cuts off the pebble flux once the growth front has reached the gap location, thus reducing the delivery of molecules to the inner disc.
Except for the largest $R_{\rm gap}$ cases, this happens before the peak enhancements of \ce{H2O} and \ce{CO2} expected in a smooth disc, which lowers the maximum enhancement and shifts the peak earlier. Generally speaking, the closer in the trap, the sooner it interrupts the dust flux, and the stronger these effects are, as was shown for \ce{H2O} by \citet{Kalyaan_2021,Kalyaan_2023}. 
Moreover, as \ce{H2O} is depleted more quickly by advection than \ce{CO2}, it reacts faster to the loss of supply. As a result, traps act to increase the \ce{CO2}/\ce{H2O} vapour mass ratio, which now peaks in the range $30-300$, compared to at around $10$ for a smooth disc. This is consistent with \citet{Mah_2024} where sufficiently deep gaps raised the inner disc C/O.

At late times, the models with traps produce vapour masses that flatten off. While smooth discs are constantly losing both gas and dust to the star, discs with traps by definition are keeping hold of a much larger reservoir of dust, which slowly leaks through the traps and thus sustains a minimum supply of vapour mass to the inner disc. We explore this more in Sect. \ref{sec:trapTrends}. Generally the dust that leaks grows inefficiently and remains small and well coupled. Hence in this phase, there is no difference between the transport of vapour and ices, so \ce{CO2} and \ce{H2O} are delivered at the same net rate, fixed by their ice abundances, and thus their ratio tends back towards its initial value.

One special case is where the trap is sited at $1\,\mathrm{au}$, which is outside the \ce{H2O} snow line but inside that of the \ce{CO2}. In this case, it has little effect on the \ce{CO2} vapour evolution but a very significant one on the \ce{H2O} vapour evolution. This results in a much earlier rise in the \ce{CO2}/\ce{H2O} vapour mass ratio, which reaches $\sim 1000$ on timescales $\sim0.3\,\mathrm{Myr}$, higher than achieved in any of the other models by a factor of a few.
The 1 au case also shows, uniquely, a resurgence of the \ce{H2O} vapour mass due to leaking dust. This happens largely because the dust mass is still accumulating in the trap on million year (Myr) timescales. This long-term supply of \ce{H2O} leaking from the trap is similar to that found for traps of intermediate depth by \citet[][their so-called `traffic jam scenario']{Mah_2024} which balance the ability of deep traps to retain some dust with the ability of shallow traps to then leak that dust. 
However, because all the \ce{CO2} enters the gas phase before the pebbles are trapped, the dust in the trap is \ce{CO2}-poor. Thus, the slow leaking of dust from the trap is not able to replenish the \ce{CO2} vapour, resulting in very low \ce{CO2}/\ce{H2O} ratios on timescales $\gtrsim3\,\mathrm{Myr}$.

The column densities tell a similar story. Despite the reduced dust flux in the presence of traps, the inner disc never becomes sufficiently optically thin to deviate from Eq.\,\ref{eq:expectedNvis}, and thus in all cases the \ce{H2O} column density is fairly constant in time, varying by no more than factor $\sim3$, considerably less than the variation in the underlying \ce{H2O} vapour mass.
On the other hand, despite the fact that traps reduce the delivery of \ce{CO2} to the inner disc, by blocking dust from reaching the inner disc they also allow the visible column density to undergo its rapid rise much sooner and thus tend to promote higher retrieved \ce{CO2} column densities.
Moreover, as the dust obscuration affects both the \ce{H2O} and \ce{CO2} in a correlated way, when the \ce{CO2}/\ce{H2O} column density ratio is considered, the evolution (for any gap location) closely mirrors that of the vapour mass ratio.

With the caveat that our spectra lack contribution from the hot surface layers (see discussion in Sec. \ref{sec:methodsSpectrum}), we collect a sample of discs with so far published or submitted analysis of their MIRI-MRS spectra (Table\,\ref{tab:observed}) where slab model fits for both \ce{H2O} and \ce{CO2} are available. We limit ourselves to discs around stars $>0.3\,M_{\odot}$ because lower mass stars are known to host discs with a more hydrocarbon-rich, \ce{H2O}-poor, chemistry, which may be connected to the fact that such discs are cooler and their snow lines lie closer to the star, and hence the timescales on which their chemistry evolves due to transport are much shorter \citep{Mah_2023} than for our fiducial $1\,M_{\rm sun}$ case.
Strictly speaking, given that we do not model the hot surface layer of the disc or the extended snow surfaces, we should attempt only to compare to the warm ($250-550\,\mathrm{K}$) \ce{H2O} and \ce{CO2} components.
However given the already small sample this leaves us, and given that the works all differ in the wavelength ranges and number of components they fit for the water (and in any case only ever fit a single \ce{CO2} component that may or may not be within the temperature ranges we find) we simply plot an unweighted average over all of the \ce{H2O} fits in each work, and indicate their range with the error bar. By so doing, we make the assumption that the chemistry of all the components is correlated, for example that \ce{H2O} enhancement deep in the disc would also enhance \ce{H2O} in the surface layers (rather than being completely reset by reactions), and so that all components contain information about the bulk chemical composition even if they do not trace the bulk material directly.

For both \ce{H2O} and \ce{CO2}, we see that our models generally predict column densities at the upper limit of what is observed.
As discussed, in the case of \ce{H2O}, the value depends simply on the initial (\ce{H2O}) ice-to-dust ratio and the opacity; the \ce{CO2} column density will also be affected by equivalent quantities. Lowering the initial ice abundances, as might be more consistent with protostellar estimates, and/or raising the dust opacity, could potentially account for some of the difference between the models and the bulk of the observations. Alternatively, as the dominant line features in the slabs are often very optically thick \citep[as noted for DR Tau by][]{Temmink_2024b}, the observations may instead only trace the column down to where these features - rather than the continuum - become optically thick.
For \ce{CO2}, better agreement is seen between the models and the more \ce{CO2}-rich discs, particularly when the \ce{^13CO2} isotopologue estimates are used. This may reflect the fact that \ce{^13CO2} components are colder \citep{Vlasblom_2024b} and may potentially therefore originate deeper within the disc \citep{Bosman_2017,Bosman_2018a,Bosman_2022b}, consistent with their expected lower (line) optical depth. They may therefore be more reflective of the bulk material that we try to model rather than the warm surface chemistry. Indeed the observed temperatures of these components are closer to those that we model.
Finally, the predicted column density ratios are much closer to those observed, suggesting that the retrieved column densities of the two molecules may be less than our predictions due to similar effects.

Focussing therefore on ratios, firstly for two discs with \ce{^13CO2} detections - GW Lup \citep{Grant_2023} and CX Tau \citep{Vlasblom_2024b} - the former is most consistent with a \ce{CO2}/\ce{H2O} column density ratio predicted by a model with a gap at large radii - consistent with the presence of a single known gap at 74 au \citep{Huang_2018} - while the latter is more consistent with a gap closer than 7 au, or a completely smooth disc - consistent with its status as a compact, `drift-dominated', disc \citep{Facchini_2019} with no resolved substructures down to 5 au.
The water-dominated DR Tau \citep{Temmink_2024a,Temmink_2024b} is a compact disc, with no substructures that are clearly resolved in the image plane. Our models tentatively suggest that it is most consistent with a gap very close to the star, between the \ce{H2O} and \ce{CO2} snow lines, that is, leaking considerable amounts of \ce{H2O} on Myr timescales.
The other discs are all known to be substructured, though AS 209 has a much more complicated system of multiple rings \citep{Huang_2018} which may be why it agrees the least, and SY Cha is a transition disc with a large millimetre dust cavity \citep{Schwarz_2024}.
Overall, this may suggest that \ce{CO2}-dominated spectra reflect discs where transport controls the chemistry while other spectra represent scenarios where chemistry is determined by other processes such as chemical processing in the warm molecular layer. Alternatively, the lack of clearly detected hot bands in most of these discs (due to strong blended \ce{H2O} emission) may bias the observations away from high column densities. \citet{Grant_2023,Gasman_2023} show that the Q-branch alone can be fit with an optically thin, low column density solution, but that including the hot bands leads to column densities more than an order of magnitude higher in which the Q-branch is optically thick and insensitive to column; this would explain why Sz 98, where those bands are tentatively detected, shows a column density more similar to the discs with \ce{CO2}-dominated spectra and \ce{^13CO2} detections.

We find that, especially in the case of water, the effective emitting radius and temperature derived from the slab fits are mostly determined by the location of the snow line and the average disc temperature interior and show almost no evolution (e.g. Fig. \ref{fig:4vs3}). Therefore, we do not include plots of $T$ or $R_{\rm eff}$ for our full grid of models.

\subsubsection{Scenario 2: Rapid drift and slow gas accretion}
In Scenario 2, there is no traffic jam inside of the snow line and therefore no inevitable hiding of the water. Thus, once the dust flux drops, the dust inside the snow line also drifts onto the star, and the delivered water is rapidly revealed. At the same time, the gas is no longer being so effectively advected onto the star, and so the vapour masses build up much more, peaking later. The result of the loss of opacity, revealing deeper layers of the disc, is an almost vertical climb in the column densities within 1 Myr to values $N_{\rm H_2O}\gtrsim10^{24}\,\mathrm{cm^{-2}}$ and $N_{\rm CO_2}\gtrsim10^{22}\,\mathrm{cm^{-2}}$, far in excess of the observed ranges for discs at $\sim1$ Myr.
The slower gas evolution means that the general pattern undergone by the \ce{CO2}/\ce{H2O} ratio (in both vapour mass and column density) of an initial fall and subsequent increase (once \ce{H2O} is accreted) happens much later (but otherwise looks similar).
Therefore, by 1-3 Myr, all models are still recovering from the initial drop in the \ce{H2O}/\ce{CO2} ratio as there has been less removal of the water by accretion onto the star. Thus, the ratio of the retrieved columns is brought much closer to the low values observed for \ce{CO2}-poor discs around AS 209 and SY Cha. However, the high absolute columns, and the inability to replicate the more \ce{CO2}-rich sources, suggest this scenario is unlikely to be the correct explanation without invoking very efficient chemical reprocessing and low resulting molecular abundances.
 
\subsubsection{Scenario 3: Rapid drift and faster gas accretion}

Much like in Scenario 2, in Scenario 3 there is no traffic jam inside the water snow line and therefore the column densities increase sharply after the pebble flux is interrupted by the trap. However the assumptions of different $\alpha$ values for turbulence and viscosity means that the vapour mass enhancements do not build up quite so high or last quite so long, more like Scenario 1. The result is somewhat more reasonable column densities of $10^{20}-10^{21}\,\mathrm{cm^{-2}}$ on Myr timescales, though these are still a bit higher than observed values, similarly to the first scenario. Overall, this scenario and the first perform similarly well at reproducing the \ce{CO2}/\ce{H2O} ratio in $\sim 1\,\mathrm{Myr}$ discs, and we can draw the same conclusions about the preferred locations of any gap amongst our models. However, in this scenario there is more sensitivity to variations in the \ce{H2O} abundance than in Scenario 1.

\subsection{Trends with gap location}
\label{sec:trapTrends}
As could be seen above, the location of any gap in the disc affects the evolution of the vapour masses and column densities of different molecules through its effects on the time that the pebble flux is interrupted, the amount of pebbles that are thus isolated from the inner disc, and the amount of this isolated material that leaks through the trap. We also suggested that between two \ce{CO2}-dominated sources observed with JWST, GW Lup seems to prefer a more distant gap location and CX Tau a closer gap location (or none at all). In this section we therefore explore in more detail the relationship between gap location and the vapour masses or column densities, focusing on Scenario 1 which produced the most reasonable column densities (Scenarios 2 and 3 can be found in Appendix \ref{appendix:location}).
Figure \ref{fig:scenario1trends} shows that for the vapour masses of each molecule, there are clear trends in the vapour masses with radius.
These can be divided broadly into positive relationships with $R_{\rm gap}$ at early times in the disc evolution and negative relationships at late times.

\subsubsection{Trends in vapour mass at early times}
The positive relationship between the vapour mass and gap location at early times $\lesssim 0.5 \,\mathrm{Myr}$, which is approximately linear, can be easily understood as tracing the amount of mass that initially lies inside of the trap and is thus transported to the inner disc unimpeded:
\begin{align}
    M_{\rm d, in} &= M_{\rm d, tot} \left(1-\exp\left(-\frac{R_{\rm gap}}{R_{\rm C}}\right)\right)
    \label{eq:Mdin} \\
            &\approx M_{\rm d, tot} \frac{R_{\rm gap}}{R_{\rm C}} & (R_{\rm gap} \ll R_{\rm C}) \nonumber
            .
\end{align}
The dependence predicted by Eq.\,\ref{eq:Mdin} is plotted in Fig.\,\ref{fig:scenario1trends} (dashed-dotted line) and is indeed roughly consistent with the models, especially for \ce{CO2}.
However, the \ce{H2O} vapour delivered to the inner disc accretes on timescales $\sim0.1\,\mathrm{Myr}$, several of which have passed by the first snapshot at $0.5\,\mathrm{Myr}$. Hence in these cases, the relationship is already evolving towards its late time behaviour, making it deviate more significantly from the approximately linear trend, though it nevertheless retains a positive trend up until a time of $\sim1\,\mathrm{Myr}$.
Thus, the closer to the star the gap is opened, the less dust there is inside of it to drift, and thus we get less enhancement of the vapour masses, as noted for mm-sized pebbles by \citet{Kalyaan_2021}.

\subsubsection{Trends in vapour mass at late times}
Once the molecular reservoir that was originally present inside the gap has all been accreted onto the star, both the \ce{CO2} and \ce{H2O} vapour masses transition to a negative dependence on $R_{\rm gap}$.
At these late times, as mentioned briefly in terms of the flat vapour mass evolution, the vapour in the inner disc is supplied by whatever small dust manages to leak through the trap. Traps nearer to the star are leakier \citep{Kalyaan_2023} and so more dust is escaping.

Although small dust dynamically mostly follows the gas, it can also be produced by the fragmentation of the large grains that are tightly concentrated in the trap. Although we do not model this directly, the fixed $f_{\rm small}$ means that we will also have a population of small dust that follows the concentration of the large dust grains. This results in diffusion being a strong driver of small dust leaking through the trap. Using Eq.\,\ref{eq:Fleak}, we can estimate a diffusive flux of
\begin{align}
    F_{\rm leak} &\approx 2 \pi R_{\rm gap} \frac{\nu}{S\!c} \Sigma \frac{\epsilon_{\rm small}}{w_{\rm d}} \\
                 &= \frac{2\dot{M}_{\rm acc} f_{\rm small}}{3S\!c}  \frac{R_{\rm gap}}{w_{\rm d}} \frac{\Sigma_{\rm d}}{\Sigma_{\rm g}}\Bigg|_{\rm gap}
                ,
\end{align}
where $\epsilon_{\rm small} = f_{\rm small} \epsilon$ is the mass fraction in small dust for $\epsilon$ the total dust mass fraction, and where we use the steady state relationship $\dot{M}_{\rm acc} = 3\pi\nu\Sigma_G$. $w_{\rm d}$ is the radial width of the concentrated dust in the trap (set by the radial width of the fragmenting large grains), which can be related to the width of the pressure bump in gas by $w_{\rm d}^2 = \frac{\alpha}{S\!t} w_{\rm g}^2$ \citep[e.g.][]{Rosotti_2020}; we shall further assume $w_{\rm g} \approx w_{\rm gap}$.
For the dust-to-gas ratio, we consider $M_{\rm d, out}=M_{\rm d, tot}-M_{\rm d, in}$ (Eq.\,\ref{eq:Mdin}) to be concentrated in a ring of area $2 \pi R_{\rm gap} w_{\rm d}$, which gives $\frac{\Sigma_{\rm d}}{\Sigma_{\rm g}}\Bigg|_{\rm gap} \approx \epsilon_0 \frac{R_C}{w_{\rm d}} \exp\left(-\frac{R_{\rm gap}}{R_{\rm C}}\right)$ where $\epsilon_0 \approx 0.01$ is the initial dust mass fraction. 

The growth of large grains in the trap will be limited by turbulent fragmentation, in which case Eq.\,\ref{eq:Stfrag} predicts that $\frac{\alpha}{S\!t}\propto c_{\rm S}^2 \propto R_{\rm gap}^{1/2}$,.
Combining all of the above gives
\begin{align}
    F_{\rm leak} &\propto \frac{R_{\rm gap}}{w_{\rm d}^2} \exp\left(-\frac{R_{\rm gap}}{R_{\rm C}}\right) \\
                 &\propto \frac{R_{\rm gap}^{3/2}}{w_{\rm gap}^2} \exp\left(-\frac{R_{\rm gap}}{R_{\rm C}}\right),
    \label{eq:Fleak1}
\end{align}
where we assume $f_{\rm small}$ to be constant as per the two-population model \citep{Birnstiel_2012}\footnote{However, we note that, in reality, the strong diffusive gradient will assist some of the marginally decoupled dust to leak out of the trap, making the fractional population of dust susceptible to leaking somewhat higher than $f_{\rm small}$. Only a full fragmentation-coagulation treatment \citep[e.g.][]{Stammler_2022} could capture this behaviour accurately, but implementing molecular tracers in such a model is highly non-trivial.}.  

\begin{figure}[t]
    \centering
    \includegraphics[width=\linewidth]{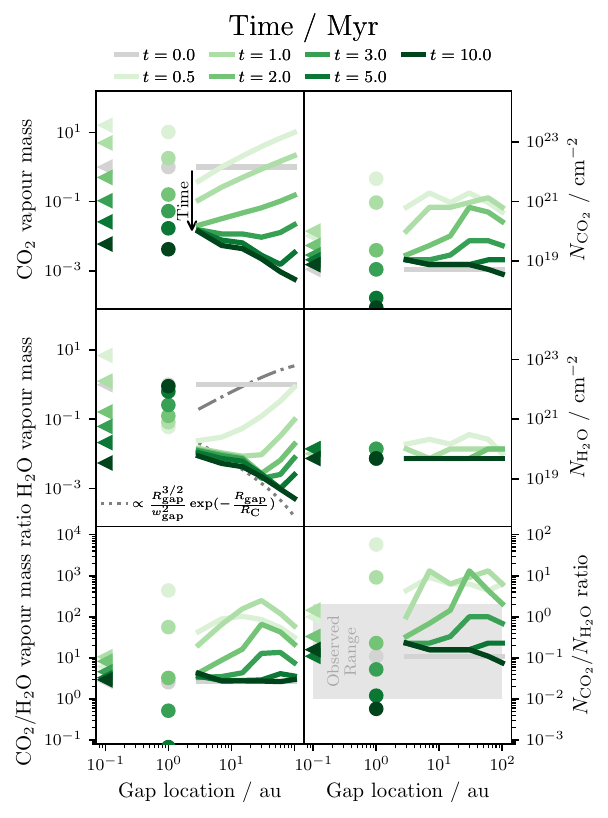}
    \caption{Trends of \ce{CO2} (top) and \ce{H2O} (middle) vapour masses (left) and column densities (right) and their ratio (bottom) with $R_{\rm gap}$ at different stages of evolution for discs with traffic jams inside the water snow line (Scenario 1). The initial condition is in grey, while increasingly dark shades of green correspond to increasingly late times. The model with $R_{\rm gap}=1\,\mathrm{au}$ is plotted separately at the dots as its gap is located very close to the \ce{H2O}{} snow line and thus inside the \ce{CO2}{} snow line. A smooth disc model is included for comparison as the triangles. Equations \ref{eq:Mdin} (dashed-dotted) and \ref{eq:Fleak} (dotted) are plotted as guides to the eye the trends for approximate trends at early and late times.}
    \label{fig:scenario1trends}
\end{figure}

Therefore, since our models assume $w_{\rm gap} \propto H_{\rm gap} \propto R_{\rm gap}^{5/4}$, then this gives $F_{\rm leak} \propto R^{-1}$ for $R_{\rm gap} \ll R_{\rm C}$ when turbulent fragmentation limits dust growth throughout the trap.
The limit from Eq.\,\ref{eq:Fleak1} is plotted as the dotted line in Fig.\,\ref{fig:scenario1trends}
While the agreement is imperfect as a) we made several approximations in the calculation of the dust flux and b) the two population model does not completely accurately capture the nuances of diffusive flux, there is sufficient agreement with the general negative trend to be confident that diffusion of dust from the trap is the explanation for the trend.

In conclusion, the negative trend with $R_{\rm gap}$ results principally from the trapped dust being more tightly concentrated in traps closer to the star, which induces a stronger diffusive gradient.

\subsubsection{Observable trends}
As before, the \ce{H2O} vapour evolution leads that of the \ce{CO2}.
This can be seen most clearly on timescales of a couple of Myr, where \ce{H2O} has adopted a mostly negative relationship with $R_{\rm gap}$ (save for the cases with the most distant gaps where the most volatiles are delivered and consequently it takes the longest to deplete them to the same level) while the \ce{CO2} still has a positive trend. This is important because it means that the \ce{CO2}/\ce{H2O} ratio is generally an increasing function of $R_{\rm gap}$ (though declines again slightly for $R_{\rm gap}\gtrsim 60\,\mathrm{au}$ as the gap location becomes close to the disc outer edge at $R_{\rm C}=70\,\mathrm{au}$) on timescales of 1-3 Myr. However, at late times, the ratio is generally independent of trap radius (except for if the trap lies between the two snow lines) and is simply determined by the input ice abundance ratio because the supply of molecules is controlled by leaking from the traps and hence both \ce{H2O} and \ce{CO2} leak at the same rate.

If measurable, the existence of distinctly opposite trends at the early and late stages of disc evolution would be a powerful one. It would potentially allow us to measure the age of a population of structured discs and their traps relative to the timescales of the dust and gas evolution that control the trends. Therefore, we finally assess whether such trends carry over to the column densities of the molecules retrieved from slab fits.

As expected (given the previously discussed insensitivity of the column density to the vapour mass in Scenario 1), the trends with respect to \ce{H2O} are lost. 
Conversely, we generally recover the positive relationship expected between the \ce{CO2} column density and gap radius at early times, as there is no dust pile up between the \ce{CO2} and \ce{H2O} snow lines, giving us sensitivity to the effect of close in traps reducing the enhancement of \ce{CO2} by dust delivery. However, at late times, we see the same column densities emerging for all traps, despite the fact that the closer in traps are leakier. This is because the leakier traps are supplying more small dust - all of which remains coupled to the gas ---which contributes to increasing the continuum optical depth and obscuring the delivered \ce{CO2} (much like applies inside the \ce{H2O} snow line at all times due to the traffic jam).
Finally, because the \ce{H2O} column density never shows any trend, these trends for the \ce{CO2} column density carry over to the ratio of the \ce{CO2} and \ce{H2O} column densities. Given that dust optical depth ultimately affects both molecules similarly, the trends in column density ratio reflect the underlying \ce{CO2}/\ce{H2O} vapour mass ratio relatively well.

\begin{figure}[th]
    \centering
    \includegraphics[width=\linewidth]{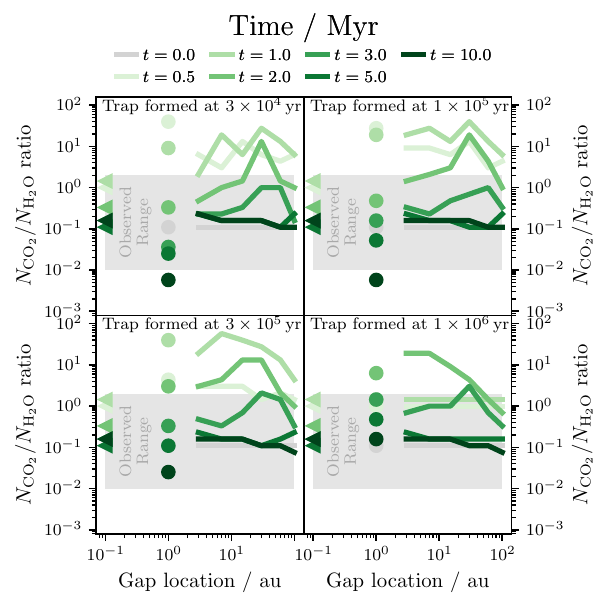}
    \caption{Dependence of the \ce{CO2}/\ce{H2O} column density ratio on gap location at different snapshots in the disc evolution. The panels show models in which the traps are fully developed at successively later times.}
    \label{fig:traptiming}
\end{figure}
\subsubsection{Dependence on trap formation time}
Having identified the \ce{CO2}/\ce{H2O} column density ratio as a reasonably promising tracer of the role of drift and traps preventing drift, we now consider the effect of traps forming later (up to 1 Myr into the disc's evolution) on its predicted trends with $R_{\rm gap}$.

The time taken for dust to start drifting is generally an increasing function of radius, as the dust is assumed to start small and need some time to grow, with the growth timescale which delays the onset of drift scaling as $1/\Omega_{\rm K} \propto R^{3/2}$ \citep{Birnstiel_2012}. As long as a trap forms before the dust at that radius (and exterior) starts to drift, then it will have the same effect as a trap formed at $t=0$. However, if the dust has already started to drift past that radius once the trap forms, then the amount of dust it can prevent reaching the inner disc is reduced. Combining these considerations, we can see that delaying the onset of trap formation will have the strongest effect on traps closer to the star where dust starts to drift sooner.

The trend at early times is controlled by the amount of dust that is not blocked by the trap, which is reduced most by close-in traps. Later trap formation therefore allows more dust into the inner disc before close-in traps block it, leading to more enhancement of the inner disc; the outer traps are not so affected because most dust lies interior to them anyway. The result, as illustrated by Fig.\,\ref{fig:traptiming}, is to make all gaps act more like the gaps further out in terms of their impact on \ce{CO2}/\ce{H2O}. This makes the positive trends with $R_{\rm gap}$ shallower: if the traps still form within $\lesssim 0.1\,\mathrm{Myr}$ (the rough growth+drift timescale for the outer disc) a majority of the dust will still be outside any trap once it opens, and a positive, albeit less strong, relationship with $R_{\rm gap}$ on $\sim 2$ Myr timescales is still maintained. On the other hand, such trends are either largely indiscernible or even reverse if the traps form after $\gtrsim0.1\,\mathrm{Myr}$. 
Likewise, \citet{Mah_2024} found 0.1 Myr to be a critical timescale for trap formation to prevent significant \ce{H2O} enhancement, and showed that otherwise the enhancement would last the whole disc lifetime for $\alpha\lesssim10^{-4}$.

There are known examples of rings suggestive of dust traps in Class 0/I discs at young ages \citep{ALMApartnership_2015,Segura-Cox_2020}, but such cases are rare, with surveys suggesting only 10\% incidence for substructures $\gtrsim10\,\mathrm{au}$ \citep{Ohashi_2023,Hsieh_2024}.
If any connection can be demonstrated between dust substructure and inner disc chemistry, this would imply a significant population of unresolved or unobservable substructures in these discs, potentially due to the dust emission being both optically and geometrically thick \citep{Ohashi_2023,Sharma_2023,Guerra-Alvarado_2024}.
This would be consistent with the population synthesis modelling of \citet{Delussu_2024} which requires substructures capable of trapping dust to be nearly ubiquitous and to form by $0.4\mathrm{Myr}$, in order to match the millimetre spectral indices, fluxes, and sizes of discs.
However, it is hard for planets to grow large enough to open gaps at such times, especially given that younger discs are warmer; even with the generous assumption that planetary embryos can form in the Class 0 phase, less than 10\% of discs are likely massive enough to produce gap-opening planets in under 0.1 Myr \citep{Nazari_2024e}.
Indeed, although the relative mass budgets of Class II discs compared to planetary systems \citep{Manara_2018} and embedded discs \citep{Tychoniec_2020} point towards the solids that become planets being locked up into larger bodies before the Class II stage, this may only have time to progress as far as planetesimals in the embedded phase \citep{Cridland_2022}.
Consequently, a strong connection between dust substructure and inner disc chemistry would likely require a non-planetary origin for the substructures such as magnetohydrodynamic zonal flows \citep[see discussion in][]{Bae_2023}, or instabilities related to magnetohydrodynamic winds \citep{Riols_2019}.

\section{Discussion}
\label{sec:discussion}

\subsection{What causes \ce{CO2}-dominated spectra?}
Our models show that the inner disc \ce{CO2}/\ce{H2O} ratio in the bulk of the disc can be traced through 0D LTE slab model fits to MIR spectra with components of $200-300\,\mathrm{K}$ under assumptions of efficient vertical mixing and negligible chemical reprocessing. This ratio is less affected by continuum opacity effects than the individual columns and therefore is potentially a better tracer of the inner disc chemical evolution as a result of gas and grain transport than the features of the individual molecules. This is because the abundance of the two molecules evolves on different timescales, leading to a predictable pattern in the ratio of their vapour masses due to drift while the relative evolution of the molecular gas and the dust is more uncertain. The column density ratio closely reflects the pattern shown in the vapour masses for a range of dust evolution scenarios.

\subsubsection{Measuring \ce{CO2}/\ce{H2O} and Limitations of 1D models}
To use the \ce{CO2}/\ce{H2O} ratio as a tracer of drift, one needs to be able to retrieve it reliably from observed spectra. Our models, at the 1-3 Myr ages of most discs characterised so far, mostly predict higher column densities than observed; the predicted temperatures are also lower than the observed temperatures.

One possible reason for this would be that the observed spectra trace a relatively thin vertical layer that is dominated by the disc's surface layers; here gas temperatures are hotter and molecules are exposed to higher UV fluxes, leading to faster chemical destruction and a reduction in molecular abundance. In other words, the disc surface has much shorter chemical timescales than estimated at the midplane in Appendix \ref{appendix:rates}, whereas our model ignores chemical reactions entirely. This would be in line with the conclusions of \citet{Bosman_2018a} that a vertically averaged destruction rate at least 1000 times higher than that at the midplane would be needed to explain the then inferred \ce{CO2} abundances.

Nevertheless, the existence of relationships between the cold \ce{H2O} emission and the disc structure \citep{Banzatti_2023,RomeroMirza_2024b} - while the hotter emission component shows no such relationship but instead correlates with the accretion rate - suggests that chemical reprocessing in the surface layers does not result in the complete destruction of a drift signal with some signature remaining in the colder emission. Even if the exact $N_{\rm\ce{CO2}}/N_{\rm\ce{H2O}}$ is altered, it should still correlate with C/O, which evolves following the delivery and accretion of \ce{H2O} and \ce{CO2} \citep{Mah_2023}. Put another way, any signature of delivered \ce{CO2} should last longer than any signature of delivered \ce{H2O}. Nevertheless, 2D thermochemical modelling is needed to determine how the delivered ratio translates into the observable ratio and make the probe more quantitatively robust.

A second difference between our modelling approach and real spectra is that we always use the entirety of the molecular spectrum within our modelled wavelength range. For example the full structure consisting of several bands and branches of \ce{CO2} emission shown in Figs. \ref{fig:exampleSpectrum}, \ref{fig:exampleSpectrumLate}, and \ref{fig:exampleSpectrumLater} may allow us to break the degeneracy between column density and temperature for optically thick features \citep{Bosman_2017,Bosman_2022b}. Conversely, in many observed spectra only the $15\,\mu\mathrm{m}$ Q-branch of \ce{CO2} is unambiguously identifiable amongst the forest of \ce{H2O} lines.

Notably, the discs that come closest to our models are those where additional \ce{CO2} features are seen. \citet{Grant_2023} and \citet{Vlasblom_2024b} show that the estimated column densities are much higher when the weaker \ce{^13CO2} Q-branch feature is used and imply the main isotopologue has become optically thick. This is consistent with modelling predictions that the main contribution to the \ce{^13CO2} emission should come from slightly deeper in the disc \citep{Bosman_2017,Bosman_2018a} and be a sensitive tracer of drift, a hypothesis that is also supported by the retrieved excitation temperatures of \ce{^13CO2} being colder than those of \ce{CO2}, more similar to our modelled temperatures.

The degeneracy between the effects of column high optical depth and high temperature on the broadening of the \ce{CO2} $15\,\mu\mathrm{m}$ Q-branch means that this feature could be fit with a hot, optically thin, solution \citep{Salyk_2011}. However, \cite{Grant_2023} demonstrate how including the hot bands lowers the retrieved temperature and raises the retrieved column density substantially. A similar effect was seen by \citet{Gasman_2023}, and \citet{Colmenares_2024} also include a hot band in their fits and find an above average $N_{\rm CO_2}/N_{\rm H_2O}$. This is consistent with the expectations from thermochemical modelling that the hot bands trace deeper emission than the $15\,\mu\mathrm{m}$ Q-branch \citep{Bosman_2022b}.
Finally, for fits to protostellar spectra, the P and R-branches alone give a much lower temperature and higher column than when the Q-branch is also included \citep[Fig. D.1;][]{vanGelder_2024}.

Overall, it appears key that weaker features sensitive to higher molecular column densities and deeper depths in the disc are used to mitigate against the retrieval being dominated by hot, chemically altered gas at the disc surface and thus estimate the bulk ratio more accurately. For \ce{CO2}, this is likely mainly possible for already quite \ce{CO2}-rich discs where these features are clear amongst the forest of other emission lines and can more easily dominate the spectrum.

\subsubsection{Interpreting \ce{CO2}/\ce{H2O}}
In smooth discs, both \ce{H2O} and \ce{CO2} are delivered efficiently by dust to the inner disc, but water is accreted onto the star faster. Therefore, the \ce{CO2}/\ce{H2O} ratio tends to be enhanced by radial drift on timescales $0.5-1\,\mathrm{Myr}$ \citep{Mah_2023} before declining back towards a baseline. Thus, even smooth, drift-dominated discs can produce \ce{CO2}-dominated spectra when observed on Myr timescales (assuming a solar mass star and $\alpha \sim 10^{-3}$), as shown by the reasonable agreement between our model and the observations of CX Tau \citep{Vlasblom_2024b} when using the estimate from the \ce{^13CO2} isotopologue. In this picture of sequential delivery as a result of drift, more \ce{H2O}-rich spectra may represent discs that are younger or undergoing slower evolution (e.g. Scenario 2). Relatively more evolved sources should show weaker \ce{H2O} emission while the \ce{CO2} has not yet declined, allowing clearer identification of weak \ce{CO2} features that typify \ce{CO2}-dominated discs.

We also found that dust trapping tends to enhance \ce{CO2}/\ce{H2O}. As the \ce{H2O} is accreted onto the star faster than \ce{CO2}, the best way to achieve this on Myr timescales is to have interrupted the delivery of \ce{H2O} early enough for it to have all accreted, but late enough to still allow plenty of \ce{CO2} to have been delivered. Consequently, traps at 10s au (which still let the early drift phase happen but do not leak much dust and ice at later times and therefore cannot replenish the accreted water) are more effective at enhancing \ce{CO2}/\ce{H2O} than traps at small radii. However, one special exception is where the trap lies between the snow lines of \ce{H2O} and \ce{CO2} in which case \ce{CO2} arrives unimpeded to the inner disc, while almost no dust can reach the water snow line.

Different discs also show variations in the temperatures and emitting areas which potentially point towards a difference in the 2D thermochemical structures of their inner regions.
One way in which the inner disc structure can be altered such as to affect the MIR spectrum is with to the presence of a deep gas cavity, unresolved in the ALMA observations \citep{Vlasblom_2024}. While small cavities can increase the emitting area due to the heated inner wall, cavities of a few AU in size reduce the area over which the disc is hot enough to desorb ices and also form molecules through gas phase reactions; this tends to enhance the \ce{CO2} line strengths relative to those of \ce{H2O}.
However, the impact on retrievable column densities was not explored, and these may be more weakly affected if the visible column density is a weak function of radius as in Fig\,\ref{fig:exampleSpectrum}. While cavities nevertheless remain a viable option, our models suggest an alternative route to producing \ce{CO2}-dominated spectra; future high-spatial-resolution submillimetre observations would help to distinguish the two scenarios.

One mechanism that may lead to the formation of cavities of a few au is internal photoevaporation, with this occurring once the accretion rate falls below the photoevaporation rate \citep{Clarke_2001}. In such a case, the inner disc is completely isolated and thus no leaking of dust into the inner disc is expected once the cavity opens. \citet{Lienert_2024} explored this scenario in a disc undergoing X-ray photoevaporation \citep{Picogna_2021} and with $\alpha=10^{-4}$ and hence slow viscous evolution (c.f. our Scenario 2). In this case, they find that photoevaporation opens a cavity before the \ce{CO2}-rich vapour reaches the inner disc (a few Myr, see Fig \ref{fig:scenario2evolution}) and hence the \ce{CO2}-dominated phase is prevented. Moreover, given the weak accretion implied by a low $\alpha$, the sublimated water is trapped in the vicinity of the snow line by the cold finger effect and is poorly accreted, leading to a resurgence of the \ce{H2O} abundance in the inner disc during the accretion of the inner disc.
Thus, whether cavities indeed lead to \ce{CO2}-dominated or \ce{H2O}-dominated spectra depends sensitively on the timing of their formation and the relative roles of equilibrium (photo)chemistry and transport.

Finally, \citet{Vlasblom_2024} also show that having more small dust present in the upper layers reduces the temperatures and therefore the size of the hot surface layer and the abundance of \ce{H2O} relative to \ce{CO2} therein. This could also lead to \ce{CO2}-dominated spectra. Such dust enrichment could also be the result of drift \citep[as suggested for IM Lup by][]{Bosman_2023}, and so this effect may broadly act in the same direction as our results. Future modelling should therefore consider if the temperature structure of the inner disc changes as a result of dust drift.

\subsection{Survival of dust upon ice sublimation}
The outcomes of our modelling, particularly the sensitivity to the enhancement of \ce{H2O} by radial drift (as well as the late time behaviour dominated by the leaking of small dust from the traps), are dependent on how well-coupled the dust grains are in the inner disc.
In the two-population model \citep{Birnstiel_2012} on which we base our dust, it is assumed that dust can grow to an equilibrium large size resulting from coagulation and - unless drift removes grains faster - collisional fragmentation. In Scenario 1, this leads to a change in the dust size across the \ce{H2O} snow line due to the assumption that dry grains are more fragile.
However such an approach neglects the fact that dust grains can, realistically, have a much more complicated internal structure: for example if they coagulate while completely covered in ice, then the sublimation of this icy glue could lead to the dust grains fragmenting. This is the `many-seeds' model \citep{Schoonenberg_2017}, which may be expected to produce well-coupled micron-sized grains \citep{Saito_2011}, much smaller than the fragmentation limit. Therefore, even in situations (e.g. Scenarios 2 \& 3) where fragmentation-limited dust grains drift, the `many-seeds' model would predict that unless the dust grains can quickly regrow to large sizes (which we see for the small dust leaking from the traps in Scenario 3 but not in Scenarios 1 or 2, Appendix \ref{appendix:location}); the outcome could be more like Scenario 1, where the dust and vapour have the same evolution and we would lack sensitivity to the evolution of the \ce{H2O}. 

In this context, \citet{Houge_2023,Houge_2024} have sought to differentiate these scenarios through modelling the collisional evolution of dust in outbursting discs, where the disc undergoes transient heating, temporarily leading to sublimation over a larger area. Although \citet{Houge_2023} argue that regrowth to equilibrium may take only several thousand years to re-establish at a few au, which is comparatively short compared to the accretion timescale, the low continuum spectral indices of the disc of the currently outbursting V883 Ori imply the presence of cm-sized grains and thus require that dust grains be indeed resilient to breakup during sublimation \citep{Houge_2024}, meaning that the many-seeds model does not apply.

\subsection{Priorities for future studies}
In order to properly test the scenarios put forward herein of the sequential delivery of volatiles to the inner disc following drift and the impact of dust traps on this picture, we suggest that observations spanning a greater range of disc ages are necessary. In our fiducial Scenario 1 at least, the period of 1-3 Myr where most data so far have been taken sees a rapid decrease in the \ce{CO2}/\ce{H2O} ratio - the best tracer that we find of the true evolution - resulting in models spanning a large range of values in line with a large scatter in the data, making it hard to discern any real evolution. Observations around the expected peak of this ratio before 1 Myr, as well as at ages older than 5 Myr would much better constrain the chemical evolution of the disc. Multiple accepted Cycle 2 JWST programmes (e.g. GO 2970-\citealt{Pascucci_2023J}, GO 3034-\citealt{Zhang_2023}) should prove useful with regards to the older discs. Moreover, observations of even younger discs would also be important to test whether discs do mostly go through a water-dominated phase earlier on.

Secondly, the comparison of models to data is made challenging by the lack of consistency as the analysis methods have rapidly developed \citep[e.g.][]{Temmink_2024b,Kaeufer_2024} since the first JWST disc spectrum published \citep{Grant_2023}. In particular, different works differ in how column densities are extracted from those that are available in terms of the number of slabs that are fit, or the spectral ranges or features to which they are fit. In order to argue for an underlying cause of a difference, for example that the higher \ce{CO2}/\ce{H2O} ratio in GW Lup than CX Tau reflects the action of a dust trap at large radius compared to that of a trap at a small, unresolved radius (or no trap at all), one needs to ensure that the values are derived consistently, and to understand better what part of the molecular budget of the disc the measurements probe. In order to refute or confirm the predicted relationships between gap location and \ce{CO2}/\ce{H2O} and therefore the role of drift and trapping in controlling inner disc chemistry, we conclude that a priority on the observational side should be to conduct a more systematic, coherent, sample analysis of JWST spectra at different phases of protoplanetary disc evolution using the weaker features that best represent the bulk material below the hot disc surface layers, while 2D thermochemical modelling needs to demonstrate the spatial origin of different spectral components and help justify the spectral analysis. Moreover, modelling should consider the influence of the evolving midplane chemistry on the more easily traced surface layers as a result of vertical transport and subsequent reactions, to allow the column density ratio to be used more quantitatively as a tracer.


\section{Conclusions}
\label{sec:conclusions}
Previous works have explored the link between dust drift and traps on the abundance of \ce{H2O} \citep{Kalyaan_2021,Kalyaan_2023} and the overall C/O ratio \citep{Mah_2023,Mah_2024} in the inner regions of protoplanetary discs. We have built on these studies in order to understand how these processes also affect the abundance of \ce{CO2} in these inner regions, and, crucially, how strongly these molecules would appear in MIR spectra depending on the disc's age and the time and location of trap formation. Our main findings can be summarised as follows:

\begin{enumerate}
      \item Ices sublimating from drifting dust as it crosses snow lines, followed by the gaseous advection of the vapour inwards, lead to the enhancement of the inner regions of protoplanetary discs with molecules. This is followed by the accretion of the molecule-rich gas onto the star. This sequence proceeds faster for less volatile molecules, leading to a \ce{H2O}-rich stage followed by a \ce{CO2}-rich stage (Fig.\,\ref{fig:vapourEvolution}).
      
      \item How well this evolution of the molecular content is traced by MIR spectra depends on the coupling of the dust grains to the gas in the inner disc, as the dust grains that deliver the volatiles provide additional continuum opacity that can hide the vapour; and this is particularly true for water. As such, there is no one-to-one relationship between column density and vapour mass. Thus, unless the amount of dust and the extinction it produces can be directly quantified, absolute column densities should be interpreted with caution, as elevated values could reflect delivered material or a loss of dust (Fig.\,\ref{fig:calibration}). Tentatively, the relatively low column densities observed with MIRI-MRS could reflect a high retention of dust in the inner disc (Scenario 1), but they could also indicate that the observed spectra are dominated by a thinner layer of hotter material at the disc surface.
      
      \item On the other hand, a ratio of column densities ---and in particular the \ce{CO2}/\ce{H2O} ratio $N_{\rm CO_2}/N_{\rm H_2O}$--- more closely reflects the underlying chemistry of the disc and can thus be used to trace the delivery of molecules to the inner disc. We predict this ratio to be enhanced on Myr timescales as a signature of this sequential drift and advection, in agreement with some observed \ce{CO2}-dominated spectra (Fig.\,\ref{fig:scenario1evolution}, grey lines), but sensitivity to weak spectral features that probe deep into the disc is likely needed to be able to see it.
      
      \item Dust traps can significantly alter the molecular abundances of the inner disc by interrupting the radial drift of dust and the supply of molecules. As a result, the peak enhancements are reduced and also occur somewhat earlier. As the \ce{H2O} vapour is accreted on a shorter timescale than the \ce{CO2}, it is more quickly affected by the lack of resupply, which increases the \ce{CO2}/\ce{H2O} ratio (Fig.\,\ref{fig:scenario1evolution}, blue lines).
      
      \item Assuming they exist from very early in the disc's evolution, closer-in traps block more dust, creating an initially positive dependence of molecular abundances on the gap location. This switches to a negative dependence, driven by the leakiness of the  traps after several viscous timescales (at the snow line). The lag of the \ce{CO2} evolution behind that of \ce{H2O} means that their ratio also has a positive dependence on gap location at $t\lesssim 2$ Myr; however, at later times, the two leak from the trap at the same rate (Fig.\,\ref{fig:scenario1trends}, bottom panels), thus erasing any dependence.

      \item Adding a delay to the formation of a trap most strongly affects close-in traps, where the growth and drift timescales are short. Formation on timescales of up to 1 Myr still prevents the long-term resupply of \ce{H2O} ice on Myr timescales, while letting more \ce{CO2} ice through initially, thus raising the \ce{CO2}/\ce{H2O} ratio. This reduces the strength of the dependence of this ratio on the trap location (Fig.\,\ref{fig:traptiming}).
      
   \end{enumerate}

\noindent
Therefore, we suggest that if carefully measured, $N_{\rm CO_2}/N_{\rm H_2O}$ retrieved from MIR spectra of protoplanetary discs can be used as a tracer of radial drift. Moreover, the presence of a positive relationship across a population of discs between $N_{\rm CO_2}/N_{\rm H_2O}$ and the location (or at least presence) of a gap could be considered as evidence of the influence of dust traps on this drift, with its strength informing us about the timescale on which the traps form. We suggest that, based on current observations of or constraints on their substructures, the \ce{CO2}-dominated discs GW Lup and CX Tau may qualitatively agree with the expected trend.

\section*{Data availability}
The version of the DiscEvolution code used for our models can be found at \url{https://github.com/AndrewSellek/DiscEvolution/releases/tag/v2.0.0}.

\begin{acknowledgements}
We thank the referee for their useful input on the comparisons between our models and different spectral features.
The authors thank members of the MINDS team - including T. Henning, I. Kamp, D. Gasman, S. Grant, N. Kurtovic, and M. Temmink - for stimulating discussions and also thank M. Temmink for advanced information about slab model fits for DR Tau.
We are also grateful to R. Booth for developing the DiscEvolution code and for discussions about implementing gaps therein.
A.D.S., M.V., and E.v.D. acknowledge support from the ERC grant 101019751 MOLDISK.
E.v.D. also acknowledges support from the Danish National Research Foundation through the Center of Excellence “InterCat” (DNRF150) and grant TOP-1 614.001.751 from the Dutch Research Council (NWO).
      
\end{acknowledgements}

%
%

\bibliographystyle{aa}
\bibliography{biblio}

%

\begin{appendix} 

\section{Transport vs chemical processing}
\begin{figure*}[ht]
    \centering
    \includegraphics[width=0.9\linewidth]{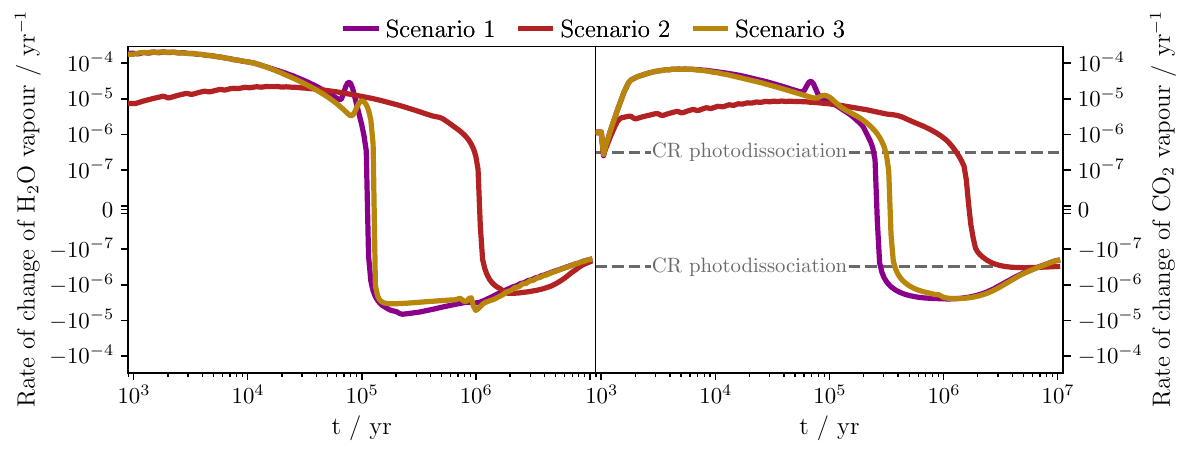}
    \caption{Evolution of the rate at which the vapour masses of \ce{H2O} (left) and \ce{CO2} (right) are changing in a smooth disc model in Scenarios 1 (purple), 2 (gold), and 3 (red) as a result of delivery by drifting dust and viscous advection onto the star. A typical rate for cosmic ray (CR) photodissociation of \ce{CO2}, $10^{-14}\,\mathrm{s^{-1}}$, is shown as the dashed lines as a reference for the rates at which chemical reactions occur.}
    \label{fig:rates}
\end{figure*}
\label{appendix:rates}
Here we set out to validate the assumption that for the main species considered in this work we can ignore chemical reactions.
Figure \ref{fig:rates} shows the rate on which the vapour mass $M_{\rm vap, X}$ is changing, defined as $\mathcal{R}=\frac{{\rm d}}{{\rm d}t} \ln(M_{\rm vap, X})$, for both \ce{H2O} and \ce{CO2} in a smooth disc for each scenario. Scenarios 1 and 3 present very similar evolution in this space as expected from the fact that their differences should mainly affect the retention of dust inside the \ce{H2O} snow line, and therefore only quantities relating to the amount of observable molecules.

The initial evolution is characterised by the growth and drift of dust and delivery of volatiles. As drift timescales for fragmentation-limited dust scale with radius $t_{\rm drift} \propto R^{1/2}$, and growth timescales scale roughly as $t_{\rm grow} \sim \frac{10}{\epsilon \Omega} \propto R^{3/2}$ (assuming roughly 10 e-foldings are needed to grow the dust from its initial size to when it starts to decouple), for the bulk of disc material at large radii, its ability to supply of volatiles is limited mainly by the growth timescale of the dust. The mass of dust that has grown to pebble size at time $t$ is then that contained within the pebble production front $R_{\rm grow}(t) \propto t^{2/3}$ and $M(<R) \sim R/R_{\rm C}$ for $R < R_{\rm C}$; the rate of change of the vapour mass is thus $\sim \frac{2}{3t}$. Rates of $10^{-5}-10^{-4}\,\mathrm{yr^{-1}}$ are seen on timescales of $10^4-10^5\,\mathrm{yr}$, consistent with this estimate.

Once the first generation of pebbles has passed through the disc, the pebble flux drops and the disc switches to draining the molecules through viscous accretion.
The rate, here is simply the inverse of the timescale given by Eq. \ref{eq:taccrete}. 
As a consequence, \ce{H2O}{} is removed at a rate of $\sim 10^{-5}\,\mathrm{yr^{-1}}$, and \ce{CO2} at a rate of $\sim 2\times 10^{-6}\,\mathrm{yr^{-1}}$. These are in good agreement with the most negative rates seen in Fig. \ref{fig:rates}. Scenario 3 reaches rates approximately 10 times smaller, consistent with its 10 times lower value of $\alpha$. 

Included for comparison is the photodissociation rate for \ce{CO2}{} due to the UV field produced by cosmic rays, which is typically around $10^{-14}\,\mathrm{s^{-1}}$ \citep{Bosman_2018a}. This is typically faster than the rate at which cosmic-ray-ionised He can destroy molecules. 
The delivery rates in the first $\sim10^5\,\mathrm{yr}$ are comfortably above these expected $\lesssim 3\times10^{-7}\,\mathrm{yr^{-1}}$ (photo-)chemical destruction rates of molecules and therefore we are justified to ignore (photo-)chemical reactions at the midplane during this period.
Similarly, the removal of molecules from the inner disc also happens much more quickly by gas accretion than by chemical destruction, at least for the first $\sim 5 \,\mathrm{Myr}$. The (photo-)chemical reactions can therefore only become relevant for \ce{H2O}{} and \ce{CO2}{} on the longest timescales of disc evolution, when the net timescale resulting from the balance of removal by accretion and delivery by drift tends towards smaller values because the molecular abundances are low enough to be efficiently replaced by ices sublimating from the small grains entrained with the gas. However, Scenario 3 is likely somewhat more susceptible to the removal of \ce{CO2} later in evolution happening via chemical reactions rather than accretion due to its low viscosity.

Overall, this suggests that indeed for most of a disc's evolution (except at low viscosities), the transport processes are faster than typical chemical processing timescales for the main molecules of interest in this work - at least in the vertically integrated sense that a 1D model such as ours explores - and thus we are justified in neglecting (photo-)chemical reactions.
This is consistent with the results of \citet{Booth_2019}, who saw that transport of molecules to the inner disc wiped out the signatures of chemical transformation for \ce{H2O} and \ce{CO2}.

We note that the impact of faster chemical processing at the disc surface (where molecules are exposed to higher UV fluxes) will depend on how the resulting reaction rates compare to the vertical mixing timescale. Assuming a turbulent model for the vertical diffusion with $\alpha_z=10^{-3}$, the inverse timescale will be $\sim 0.01\,\mathrm{yr^{-1}}$ at the $\sim0.5\,\mathrm{au}$ effective emitting radius of the emission. We refer to \citet{Semenov_2011} for a more detailed discussion of the comparison between vertical mixing timescales and the rates of different types of reaction.

\section{Trends with gap location for scenarios 2 and 3}
\label{appendix:location}

\begin{figure*}[p]
    \centering
    \includegraphics[width=0.9\linewidth]{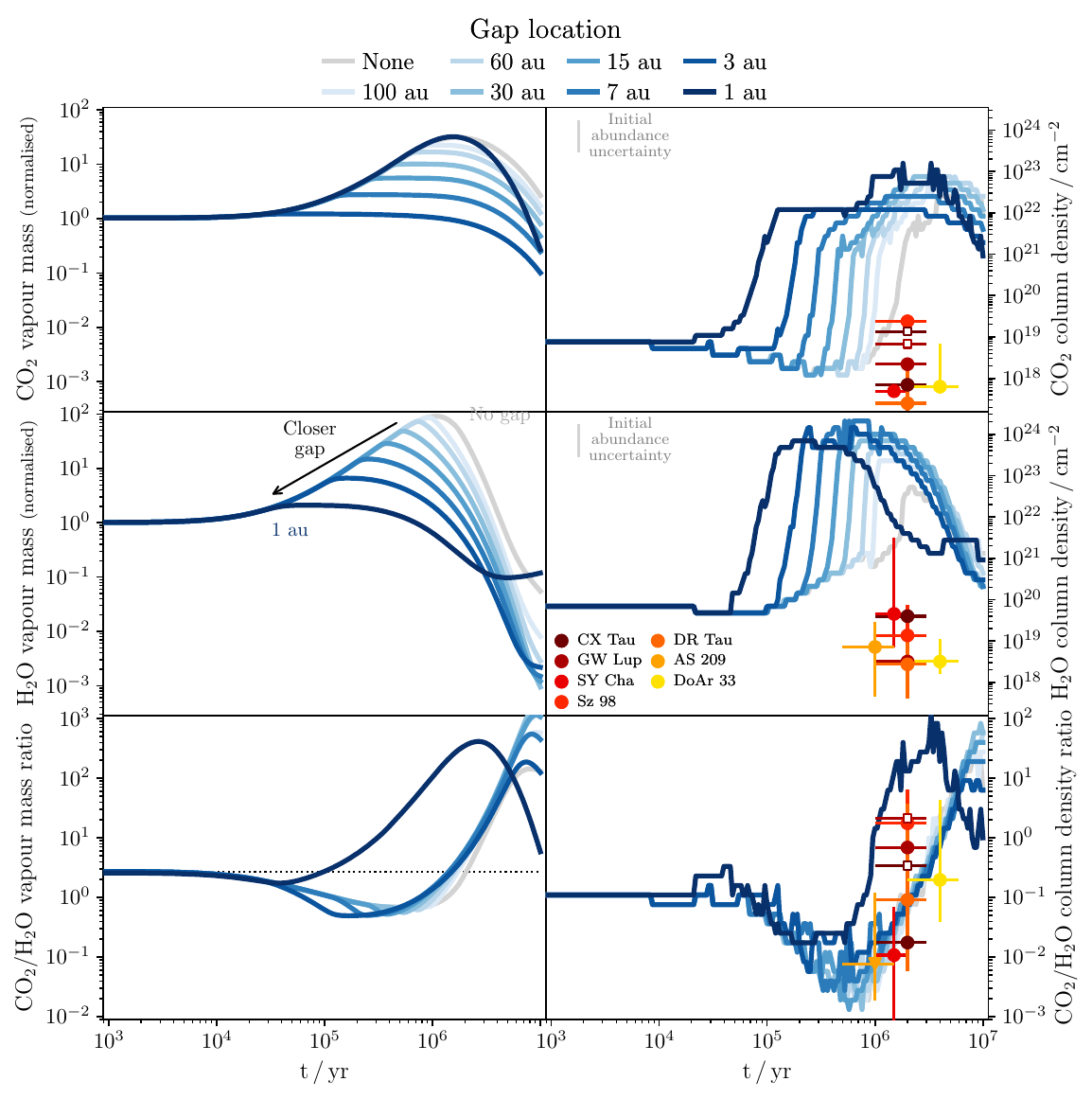}
    \caption{Same as Fig.\,\ref{fig:scenario1evolution} but for Scenario 2, which has no traffic jam inside the water snow line and a slow gas accretion.}
    \label{fig:scenario2evolution}
\end{figure*}

\begin{figure*}[p]
    \centering
    \includegraphics[width=0.9\linewidth]{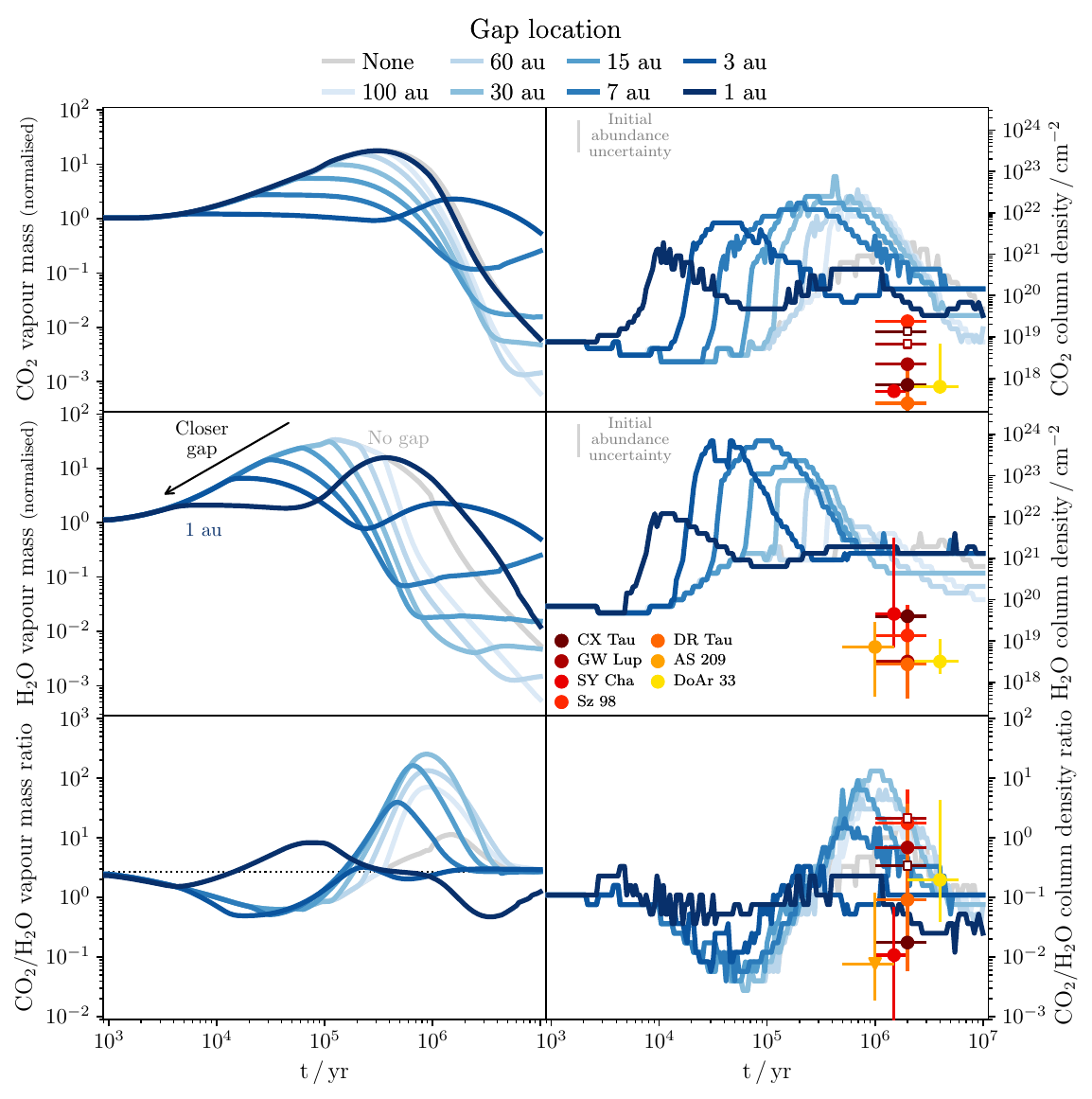}
    \caption{Same as Fig.\,\ref{fig:scenario1evolution} but for Scenario 3, which has no traffic jam inside the water snow line and a faster gas accretion.}
    \label{fig:scenario3evolution}
\end{figure*}

\begin{figure}[p]
    \centering
    \includegraphics[width=\linewidth]{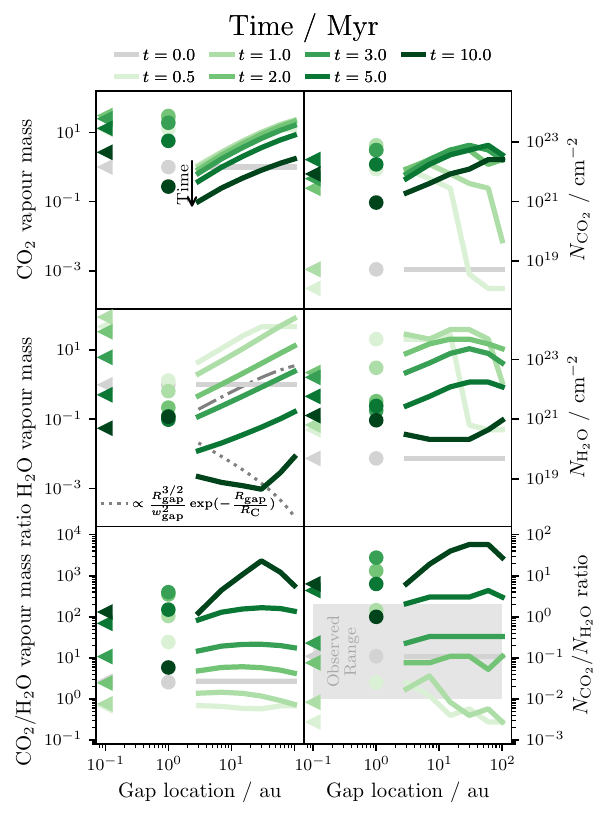}
    \caption{Same as Fig. \ref{fig:scenario1trends} but for Scenario 2, which has no traffic jam inside the water snow line and a slower gas accretion.}
    \label{fig:scenario2trends}
\end{figure}

\begin{figure}[p]
    \centering
    \includegraphics[width=\linewidth]{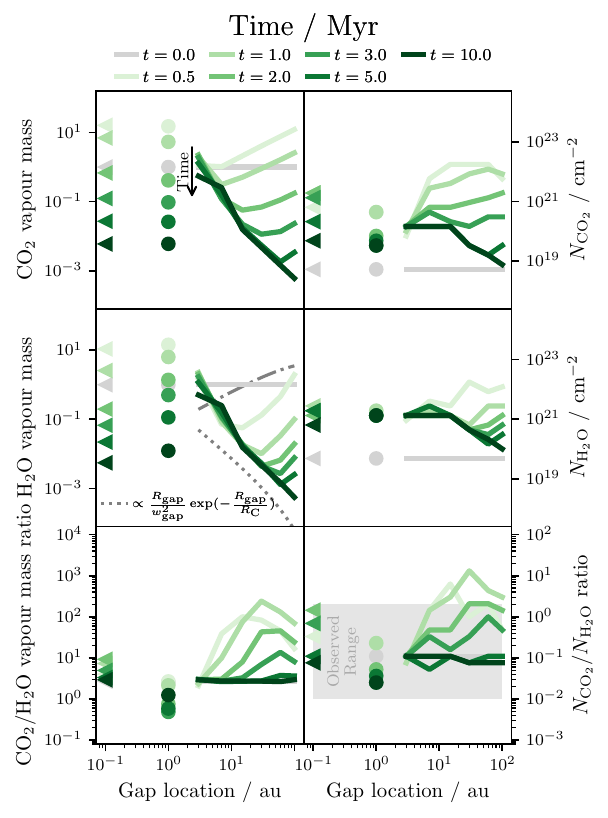}
    \caption{Same as Fig. \ref{fig:scenario1trends} but for Scenario 3, which has no traffic jam inside the water snow line and a fast gas accretion.}
    \label{fig:scenario3trends}
\end{figure}

In Scenario 3 (Fig. \ref{fig:scenario3trends}), the trends in the vapour masses strongly resemble that of Scenario 1 (Fig. \ref{fig:scenario1trends}), except for the fact the negative trend gets even steeper.
In Scenario 2 (Fig. \ref{fig:scenario2trends}) however, the evolution of the disc gas is much slower and hence the trends do not undergo a reversal during the 10 Myr modelled lifetime.
This slower evolution also means that the trends for \ce{H2O} are much closer to those predicted by Eq. \ref{eq:Mdin}.

We also note that Scenario 3 shows a steeper negative relationship than the other two. This is because in this low-turbulence scenario, fragmentation due to differential drift limits the grain sizes (apart from at the very centre of the trap). In this case, $F_{\rm leak}$ is modified to
\begin{equation}
    F_{\rm leak} \propto \frac{R_{\rm gap}}{w_{\rm gap}^2} \exp\left(-\frac{R_{\rm gap}}{R_{\rm C}}\right)
    \label{eq:Fleak2}.
\end{equation}
Since \ref{eq:Stfragdrift} predicts that $\frac{\alpha}{S\!t}\propto c_{\rm S}^2/v_{\rm K} \propto R_{\rm gap}^{0}$, then $w_{\rm dust} \propto w_{\rm gas} \propto R^{5/4}$, resulting in a steeper $F_{\rm leak} \propto R^{-3/2}$.

Moreover in this case, as one barrier to dust growth is removed, the dust can grow somewhat larger and become more concentrated, making the diffusive flux even stronger and the traps generally leakier. This is somewhat counter-intuitive but results from the assumption of a fixed $f_{\rm small}$ in the two-population model; a more realistic model of coagulation and 
fragmentation would likely suggest a smaller $f_{\rm small}$ and a net reduction in the leaking flux. Therefore, while the predicted steeper trend is probably correct, the relative values with respect to Scenario 1 may not be.
Indeed, an alternate fate for the dust trapped outside the gaps that---which would reduce the flux of dust leaking into the inner disc--- is instead to become converted to planetesimals due to the large concentration of solids. \citet{Zagaria_2023} proposed that the outer dust ring of HD HD 163296 may be undergoing planetesimal formation by the Streaming Instability (SI) and planetesimal formation has also been proposed to regulate the optical depth of dust rings, explaining the typical values observed at (sub)millimetre wavelengths \citep{Stammler_2019}.
However other works have argued that pressure bumps may not be suitable sites for SI to operate to turn millimetre-sized grains into planetesimals \citep{Carrera_2021,Carrera_2022}.
\citet{Kalyaan_2023} included planetesimal formation in their models, generally finding it only to be triggered within the traps unless the fragmentation velocities were very high, in which case some planetesimal formation also occurs at the snow line. As they focused on the earlier stages where the \ce{H2O} was enhanced in the inner disc, \citet{Kalyaan_2023} argued that these solids were trapped anyway, and hence on these timescales there were negligible differences between models with and without planetesimals. Regardless, so long as the trap is outside both the \ce{CO2} and the \ce{H2O} snow line, both ices should be equally incorporated into planetesimals and thus while the late time abundances may fall, this should not affect the \ce{CO2}/\ce{H2O} ratio. 

Unlike Scenario 1, the lack of a traffic jam inside the \ce{H2O} snow line in both Scenarios 2 \& 3 gives us more variation in \ce{H2O} column densities.
On timescales of $\gtrsim2\,\mathrm{Myr}$, we do see clear positive dependences of the column densities on $R_{\rm gap}$ in Scenario 2; however, as previously discussed, the column densities of both \ce{CO2} and \ce{H2O} are very high compared to observed values. A further consequence of such a scenario would be that any of the lines commonly used to measure \ce{H2O} emission \citep{Gasman_2023,Gasman_2025} would likely saturate and provide no sensitivity to the underlying abundances.

In Scenario 3, we actually see both the positive and negative dependences reflected in the column densities (at early and late times respectively) for both \ce{CO2} and \ce{H2O}. The ability to see the late-time trends suggests that the dust that leaks through the trap is dynamically decoupling from the gas, because the leakier traps promote a higher inner disc dust abundance and therefore more frequent collisions leading to grain regrowth in the inner disc.
In the radial-drift--limited regime, the dust surface density becomes dependent on $F_{\rm leak}^{1/2}$ \citep[c.f.][Eq.\,39]{Birnstiel_2012}, whereas the vapour surface density - or the dust surface density when the dust remains coupled - scales as $F_{\rm leak}$ \citep[c.f.][Eq.\,41]{Birnstiel_2012}. Therefore, in Scenario 3, applying the approximation of Eq.\,\ref{eq:expectedNvis}, the visible column density scales as $F_{\rm leak}^{1/2}$ thus introducing a negative dependence on $R_{\rm gap}$. For the closest-in gaps, the dust growth is so effective that it reaches the fragmentation barrier, whence it again depends linearly on $F_{\rm leak}$ and the contributions of the vapour and dust to the visible column density again cancel. Therefore, a negative relationship between the column density ratio and the location of a dust trap observed across a population of disc may indicate that dust is leaking through their trap and efficiently regrowing to drifting sizes.

\section{Additional synthetic spectra}
\subsection{Later time examples}
\label{appendix:later}
\begin{figure*}
    \centering
    \includegraphics[width=0.9\linewidth]{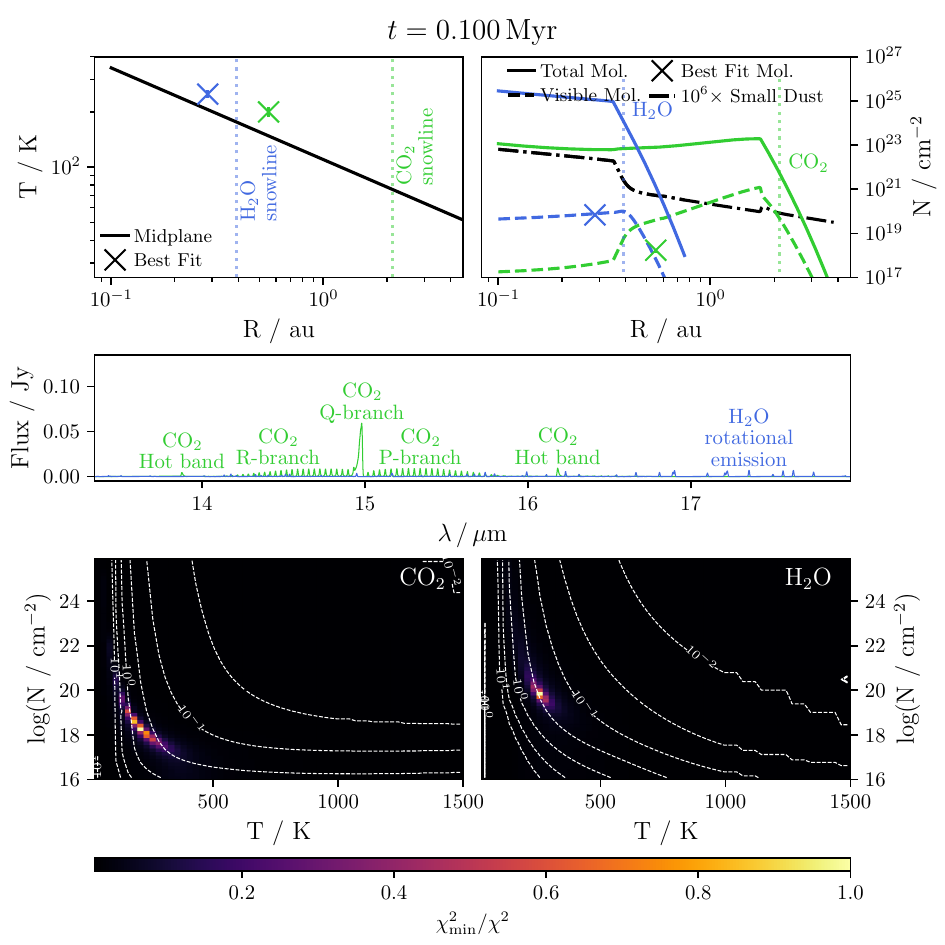}
    \caption{Synthetic spectra and retrievals as in Fig. \ref{fig:exampleSpectrum} but for a later time of 0.1\,Myr. The \ce{CO2} is overall weaker with respect to \ce{H2O} and the P- and R- branches changed shape compared to \ref{fig:exampleSpectrum}.}
    \label{fig:exampleSpectrumLate}
\end{figure*}

Figure \ref{fig:exampleSpectrumLate} shows how the input column density profiles have evolved by 0.1 Myr, when the pebble fluxes are at their peak (Fig. \ref{fig:pebbleFlux}), and the impact this has on the resulting spectrum and column density fits thereto. The \ce{H2O} column density distribution no longer shows a peak inside the snowline as the enhancement has now spread throughout the inner disc. As this is an example for Scenario 1, the same has happened for the dust (which is fully coupled to the gas inside the snowline), so the visible column density profile has changed negligibly. As a consequence, the \ce{H2O} spectrum also shows fairly little change.

Conversely, \ce{CO2} has been hidden by the build up of dust inside the \ce{H2O} snowline. The enhancement seen at its snowline has begun to spread inwards, but there is little change in the dust abundance between the snowlines. Thus the visible column density of \ce{CO2} has increased between the \ce{H2O} and \ce{CO2} snowlines. The resulting increase in emission at larger radii and decrease in emission at smaller radii biases the \ce{CO2} emission outwards, which manifests as a larger emitting radius retrieved from the slab fit. This gas is colder, which is reflected in the retrieved temperature from the slab fits and manifests most clearly in the spectrum as a shift in the peak of the P- and R- branches towards the Q-branch. Nevertheless, \ce{CO2} outside of the \ce{H2O} snowline is still quite a poor emitter and so despite the contrast in visible column density of two orders of magnitude across the \ce{H2O} snowline, there is still a significant contribution from the hotter material inside, and the resulting column density is an average of the two limits. Overall, the \ce{CO2} emission has weakened slightly compared to the \ce{H2O} at this particular time, which corresponds to a minimum in the retrieved \ce{CO2} column density (Fig. \ref{fig:scenario1evolution}).

\begin{figure*}
    \centering
    \includegraphics[width=0.9\linewidth]{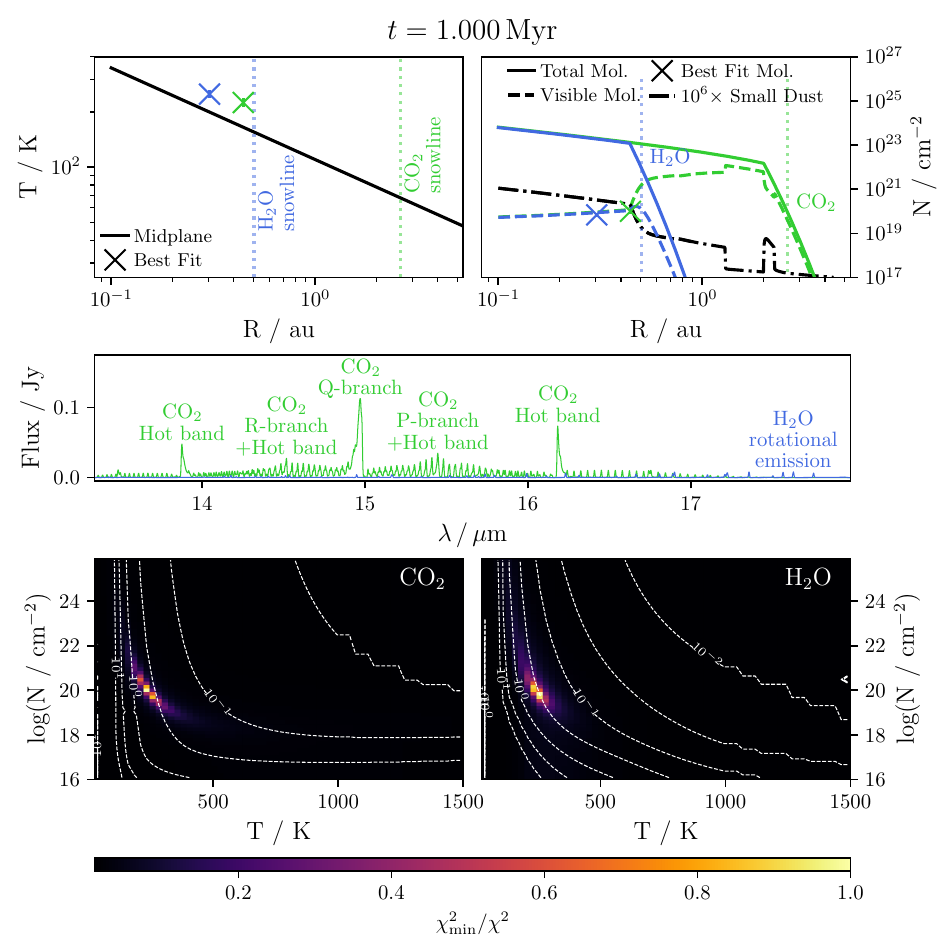}
    \caption{Synthetic spectra and retrievals as in Fig.  \ref{fig:exampleSpectrum} \&  \ref{fig:exampleSpectrumLate}. Note the different y-axis scale on the spectrum. This spectrum shows particular prominence of the hot bands over the P- and R- branches and the appearance of additional hot bands creating the bump overlaid on the P and R- branches.}
    \label{fig:exampleSpectrumLater}
\end{figure*}

For contrast we also include an example at 1\,Myr as Fig. \ref{fig:exampleSpectrumLater}. Here we see that the \ce{CO2} emission has become stronger with respect to the \ce{H2O} again, which is especially visible in the prominent `hot bands'.

\subsection{Comparison with longer wavelength emission}
\label{appendix:Ch4}
Several works have pointed to the longer wavelength \ce{H2O} emission tracing a colder component \citep[e.g.][]{Banzatti_2023,Banzatti_2024,Temmink_2024b} and found that the properties retrieved from slab fits change with the wavelength region on which the fit is performed \citep{Gasman_2023,Schwarz_2024,Temmink_2024b}. We therefore reanalysed the \ce{H2O} spectrum in the wavelength range of MIRI-MRS Channels 4A \& 4B ($17.70-24.48\,\mathrm{\mu m}$) to see if any clear differences emerged.

\begin{figure*}
    \centering
    \includegraphics[width=0.9\linewidth]{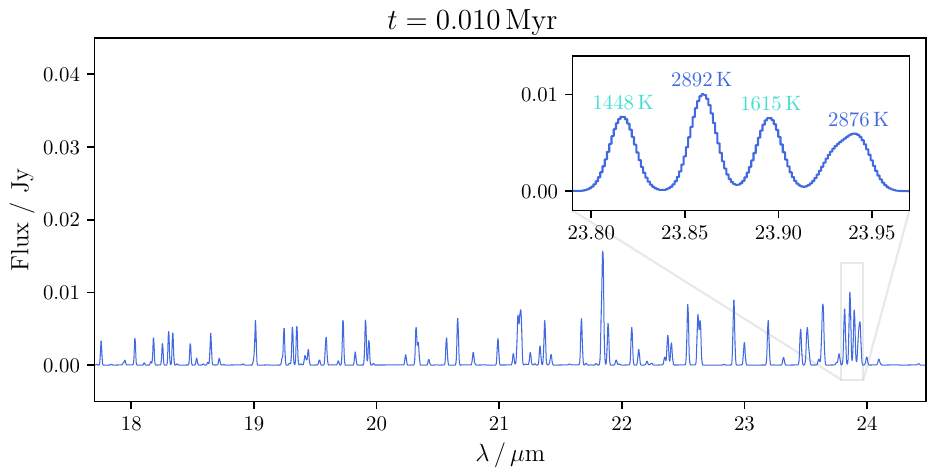}
    \caption{Synthetic spectrum of \ce{H2O} in MIRI-MRS Channel 4A \& 4B (assuming a distance of 140 pc). The inset panel shows the quartet of lines just short of $24\,\mu\mathrm{m}$ labelled with the energy of the upper level in the transition; if there were a `cold excess' \citep[c.f.][]{Banzatti_2023} the first and third lines would be stronger than the second and fourth.}
    \label{fig:spectrum_channel4}
\end{figure*}

Figure \ref{fig:spectrum_channel4} shows the spectrum in this region for the same smooth disc Scenario 1 model as Fig. \ref{fig:exampleSpectrum}, at the same time. The inset panel demonstrates that there is no excess in the lower excitation lines just short of $24\,\mu\mathrm{m}$ over their higher excitation neighbours. Such an excess has been observed in some discs \citep{Banzatti_2023,Banzatti_2024,Temmink_2024b} and is typically fit with a temperature $\lesssim180$\,K. Thus, it has typically been attributed to freshly sublimated \ce{H2O} as a result of pebble delivery. However, in our 1D model, only a narrow annulus is at these temperatures; a large emitting area - due to gas-phase \ce{H2O} in a warm, elevated, layer outside its midplane snowline - would be needed to make this component stand out but such structure is difficult to incorporate into a 1D model. 2D thermochemical modelling would be needed to determine the depth and extent of this layer and thus the resulting visible column density of water.

\begin{figure*}
    \centering
    \includegraphics[width=0.9\linewidth]{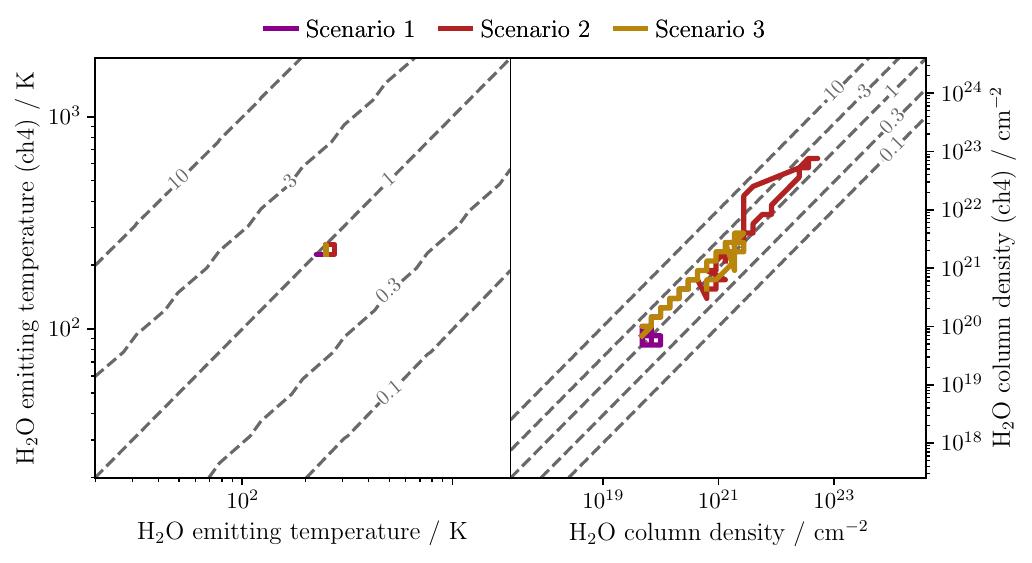}
    \caption{Comparison of the temperatures (left) and column densities retrieved from the spectrum with a single slab model for Channel 3B/C (x-axis) and Channel 4A/B (y-axis) for the three different dust evolution scenarios.}
    \label{fig:4vs3}
\end{figure*}

In Fig. \ref{fig:4vs3} we show that, likely in part due to the lack of any clear `cold excess', there is no systematic difference between the temperatures retrieved from slab model fits to this spectrum as compared to those retrieved from the shorter wavelengths (e.g. Fig. \ref{fig:exampleSpectrum}). The temperatures are, to all intents and purposes, always the same, while the column densities are essentially always within a factor of two in all dust evolution scenarios explored in this work. As a consequence, we focus on the Channel 3 values in the main body of the paper.

\end{appendix}

\end{document}